\documentclass[twocolumn,aps,prb,superscriptaddress]{revtex4-1}
\usepackage{amsmath}
\usepackage{amsfonts,amssymb}
\usepackage{graphicx}
\usepackage{epstopdf}
\usepackage{hyperref}
\usepackage{float}
\usepackage{color}
\usepackage{dsfont}
\usepackage{times}

\begin{document}
\title{Topological phases in the periodically kicked Chern insulators}
\author{Fei Yang}
\affiliation{Center for Advanced Quantum Studies, Department of Physics, Beijing Normal University, Beijing 100875, China}
\author{Zheng Wei}
\affiliation{Center for Advanced Quantum Studies, Department of Physics, Beijing Normal University, Beijing 100875, China}
\author{TianMeng Li}
\affiliation{Center for Advanced Quantum Studies, Department of Physics, Beijing Normal University, Beijing 100875, China}
\author{Su-Peng Kou}
\email{spkou@bnu.edu.cn}
\affiliation{Center for Advanced Quantum Studies, Department of Physics, Beijing Normal University, Beijing 100875, China}
\thanks{Corresponding author}
\begin{abstract}
Novel topological properties that arose in the periodically driven system are unique, in which there are two kinds of quasienergy gaps, the zero quasienergy gap and the $\pi$ quasienergy gap.
The corresponding edge modes would traverse either the zero quasienergy gap or the $\pi$ quasienergy gap, or traverse both two quasienergy gaps.
And the characterization of these two kinds of edge modes might not be the same.
However, in this paper, we find that both the zero edge modes and the $\pi$ edge modes in the Floquet Chern insulators can be characterized by the same topological invariant, where the corresponding Dirac mass term is periodically kicked.
Particularly, we take the periodically kicked Qi-Wu-Zhang model as an illustrative example. In this model, the topology is characterized by the Floquet Chern number $C_F$, and there are six different topological phases in total, denoted as $C_F=\{-1_0,-2,-1_\pi,1_\pi,2,1_0\}$.
Furthermore, we find that the Floquet operator associated with the periodically kicked Qi-Wu-Zhang model reduces to a Dirac Hamiltonian in the low-energy limit. Then, the phase diagram is uncovered by examining the topology of this Dirac Hamiltonian.
Additionally, we explore the orders of topological phase transitions in the context of Floquet stationary states by analyzing the von Neumann entropy of these states. Our work provides further insights into the topological phases in periodically driven systems.

\end{abstract}
\maketitle
\section{Introduction}
Topological phases are a class of states that have robust conducting edge or surface modes, including topological insulators (TIs) and topological superconductors \cite{review_TI_TS}.
Despite these static systems, engineering the topological phases of matter with periodic driving is also an important direction, called the Floquet system, in which the topological properties could either be analogous or beyond its static counterpart \cite{floquet_TIs_review}.
Many interesting phenomena are allowed in the periodically driven systems, such as Floquet-Anderson insulators \cite{floquet_anderson}, Floquet fractional Chern insulators \cite{floquet_fractional_chern}, anomalous chiral edge states in periodically driven system \cite{anomalous_edge_modes}, and extraordinary topological phases in non-Hermitian Floquet systems \cite{floquet_NH,floquet_NH_higher_TIs,floquet_NH_second_TIs,floquet_NH_maj,floquet_NH_topo_phase,floquet_NH_skin}.

However, different from the static systems, the topological properties of periodically driven systems are classified by their Floquet operators \cite{floquet_periodic_table,topo_cha_periodical_driven}, which corresponds to the generators of time-evolution operator over one period of driving.
The topology of Floquet operators is usually captured with the Floquet Bloch band theory \cite{floquet_Bloch,floquet_Bloch_observation}.
Since the Floquet band is unbounded, there are two inequivalent quasienergy gaps, the zero quasienergy gap and the $\pi$ quasienergy gap \cite{floquet_periodic_table}.
These gaps are generally characterized with different topological invariants \cite{anomalous_edge_modes,zero_pi_topo_num}.
However, the distinctions between the chiral edge modes traversing the zero quasienergy gap and those traversing the $\pi$ quasienergy gap are not clear yet.
Previous studies found that there is a situation that they are characterized with different winding numbers \cite{anomalous_edge_modes}. However, the reason why the Chern number is not valid for such occasion remains unclear.

In this paper, we find that when the chiral edge modes traversing the zero quasienergy gap and those traversing the $\pi$ quasienergy gap have the same chirality, then the corresponding signs of Chern number are opposite. In other words, if the chirality of these two chiral edge modes is the same, then the net value of Chern number in the Floquet system is zero. This phenomena allows for the existence of anomalous chiral edge modes \cite{anomalous_edge_modes}.
Our investigation focuses on periodically kicked Chern insulators, where the periodic driving corresponds to its Dirac mass term.
This system is particularly intriguing because its symmetry class remains identical to its static counterpart, and so is the topological invariant.
Surprisingly, we found that both the chiral edge modes traversing the zero quasienergy gap and those traversing the $\pi$ quasienergy gap can be captured with a single topological invariant, the Floquet Chern number.
We illustrate this with the periodically kicked Qi-Wu-Zhang (PK-QWZ), in which the Floquet Chern number $C_F$ is used to characterize the corresponding topological phases.
We found that there are six different topological phases in total, which corresponds to $C_F=\{-1_0,-2,-1_\pi,1_\pi,2,1_0\}$.
Furthermore, in the low-energy limit, the Floquet operator of PK-QWZ model reduces to a Dirac Hamiltonian, which shares a similar algebraic structure with its static counterpart. This insight allows us to establish the phase diagram by identifying the number of Skyrmions in the Floquet band.
Beyond the topological structure, we also investigate the implications of topological phases when the PK-QWZ model is coupled to a Markovian environment.
By appropriate design of open quantum systems, the von Neumann entropy of corresponding Floquet stationary states is studied, and the orders of phase transition are determined.

This paper is organized as follows. We briefly review the concepts of Floquet system in Sec. II.
Then, in Sec. III, we introduce the PK-QWZ model, in which the corresponding topology is characterized with Floquet Chern number, and the bulk-boundary correspondence is verified.
In Sec. IV, we study the topology contained in the Floquet band of PK-QWZ model. We identify the numbers of Skyrmions within the Floquet band and utilize this information to establish the phase diagram.
In Sec. V, the Floquet stationary states of PK-QWZ model are studied, in which we consider the situation that the system is coupled to a Markovian environment. Then the orders of topological phase transition are determined by investigating the von Neumann entropy of Floquet stationary states.
Finally, we conclude our work in Sec. VI, some future directions are discussed as well.
In the Appendices, some of the technical information is provided.
Appendix A is the expression of Floquet operator in PK-QWZ model.
Appendix B corresponds to the derivation of Sylvester equation from the Floquet Lindbladian, which is an equation of the Floquet stationary states in the periodically kicked free fermionic system.
In Appendix C, we discuss the symmetry class of periodically kicked quantum matters, where the periodic driving term corresponds to the Dirac mass term of the system.
	
\section{The periodically driven quantum systems}
The temporal evolution of a system characterized by a time-dependent Hamiltonian can be quite intricate. Nevertheless, this complexity is significantly reduced when the Hamiltonian exhibits periodic dependence on time, denoted as $H(t+T) = H(t)$.
That the long-time behavior of the system can be succinctly characterized through the utilization of a stroboscopic time evolution operator
\begin{equation}\label{time_evolution}
U(T) = \hat{\mathcal{T}}\exp\left(-i\int_{t}^{t+T}H(t')dt'\right).
\end{equation}
The Floquet state $|\psi_{\epsilon_F} \rangle$ of system only pick up a phase factor $e^{-i\epsilon_FT}$ over one complete period of driving \cite{floquet_theorem}, in other words
\begin{equation}
	U(T)|\psi_{\epsilon_F} \rangle = e^{-i\epsilon_F T}|\psi_{\epsilon_F}\rangle.
\end{equation}
From a stroboscopic perspective, $|\psi_{\epsilon_F} \rangle$ play the role of stationary states for $U(T)$, and with eigenvalues $e^{-i\epsilon_F T}$.
The factor $\epsilon_F$, called the quasienergy, is uniquely defined up to integer multiple of $\frac{2\pi}{T}$.
In other words, a solution with quasienergy $\epsilon_F$ is associated with quasienergy $\epsilon'_F=\epsilon_F\pm N\frac{2\pi}{T}$, in which $N\in\mathbb{Z}^+$.
As the result, the quasienergy is constrained within the interval $[-\frac{\pi}{T},\frac{\pi}{T})$.

In general, the stroboscopic time evolution operator $U(T)$ can be effectively modeled with $\exp\left(-iH_FT\right)$, in which the Floquet operator is formally defined as:
\begin{equation}
	H_F = \frac{i}{T}\ln\left[U(T)\right].
\end{equation}
Under this framework, the quasienergy $\epsilon_F$ corresponds to the eigenvalues of the Floquet operator $H_F$, and the Floquet states are the associated eigenvectors of $H_F$.
Amazingly, the time-independent Floquet operator is useful to grasp the topological properties of periodically driven system.
In analogy with the static case, if the quasi-band of $H_F$ exhibits nontrivial topological attributes, such as a nonzero Chern number or nonzero Winding number, then there would expect to have protected edge modes at the boundary of system.
One of the distinguishing features of Floquet systems is the unbounded nature of their quasienergy spectrum.
In particular, there are two gaps to consider between the two bands: one centered at zero quasienergy and the other centered at $\frac{\pi}{T}$.
Thus, the edge modes would traverse either the zero quasienergy gap or the $\frac{\pi}{T}$ quasienergy gap, or traverse both these gaps.

The central problem is to determine the corresponding Floquet operator, which is often intricate.
In this paper, we focus on a scenario in which the system is experiences periodic impulses, leading to a Hamiltonian of the form:
\begin{equation}\label{PK_hamiltonian}
	H(t) = H_0 + \sum_{l\in\mathbb{Z}}\delta(t/T-l)H_1.
\end{equation}
This Hamiltonian gives rise to the time evolution operator \cite{double_kicked_rotor}:
\begin{equation}\label{PK_evolution}
	U(T) = \exp\left( -iH_0T \right) \exp\left( -iH_1T \right) = \exp\left( -iH_F T \right).
\end{equation}
Then $H_F$ can be obtained by using the Baker-Campbell-Hausdorff (BCH) formula (Appendix A).
However, it's crucial to acknowledge that the eigenvalues of $H_F$, denoted as ${\lambda_F}$, may extend exceed $\frac{\pi}{T}$ or fall below $-\frac{\pi}{T}$, which is ill-defined within the Floquet system.
So, its necessary to redefine the eigenvalues $\{\lambda_F\}$.
We propose that if $\lambda_F>\frac{\pi}{T}$, then $\lambda_F$ corresponds to $\epsilon_F=\lambda_F-2N\frac{\pi}{T}$; while, if $\lambda_F<-\frac{\pi}{T}$, then $\lambda_F$ corresponds to $\epsilon_F=2N\pi+\lambda_F$, where $N\in\mathbb{Z}^+$.

\section{The periodically kicked Chern insulators}
The topology in topological insulators is characterized by the irreducible description of its Dirac mass matrix \cite{Clliford_classification}.
However, the question of how topology is modified when the Dirac mass experiences periodic perturbations remains an open issue.
For instance, let's consider the QWZ model \cite{qwz_skyrmion}, which the Bloch Hamiltonian is
\begin{equation}\label{bloch_qwz}
	H_{\text{qwz}} = \sin k_x \sigma_x + \sin k_y \sigma_y + (u_0 + \cos k_x + \cos k_y)\sigma_z.
\end{equation}
The Pauli matrices $\sigma_{x,y,z}$ represent the internal degrees of freedom in each unit cell, we assume it is the spin degrees of freedom $(\uparrow,\downarrow)$ in this paper.
The Dirac mass matrix of QWZ model is $\sigma_z$.
Then, assume that $H_0=H_{\text{qwz}}(\mathbf{k})$ and $H_1 = u\sigma_z$. By substituting $H_0$ and $H_1$ into the Eq. (\ref{PK_evolution}), one can find the Floquet operator $H_F(\mathbf{k})$ of periodically-kicked QWZ (PK-QWZ) model by using the BCH formula.
Then, how the topological properties of the system are affected by the periodic perturbation of the Dirac mass can be understood by consulting the topology in $H_F(\mathbf{k})$.

Furthermore, $H_{\text{qwz}}(\mathbf{k})$ has particle-hole symmetry
\begin{equation}
	\sigma_x H^T_{\text{qwz}}(-\mathbf{k}) \sigma_x = -H_{\text{qwz}}(\mathbf{k}),
\end{equation}
which is belongs to class D, and has $\mathbb{Z}$ classification in the 2-dimensional (2D) case.
And its topology is captured by the Chern number $C_{\text{qwz}}$, which is the integration of Berry curvature across the Brillouin zone (BZ).
Specifically, $C_{\text{qwz}}=+1$ for $0<u_0<2$, $C_{\text{qwz}}=-1$ for $-2<u_0<0$, while $C_{\text{qwz}}=0$ for $|u|>2$.

\begin{figure}[H]
	\centering
	\includegraphics[width=.5\textwidth]{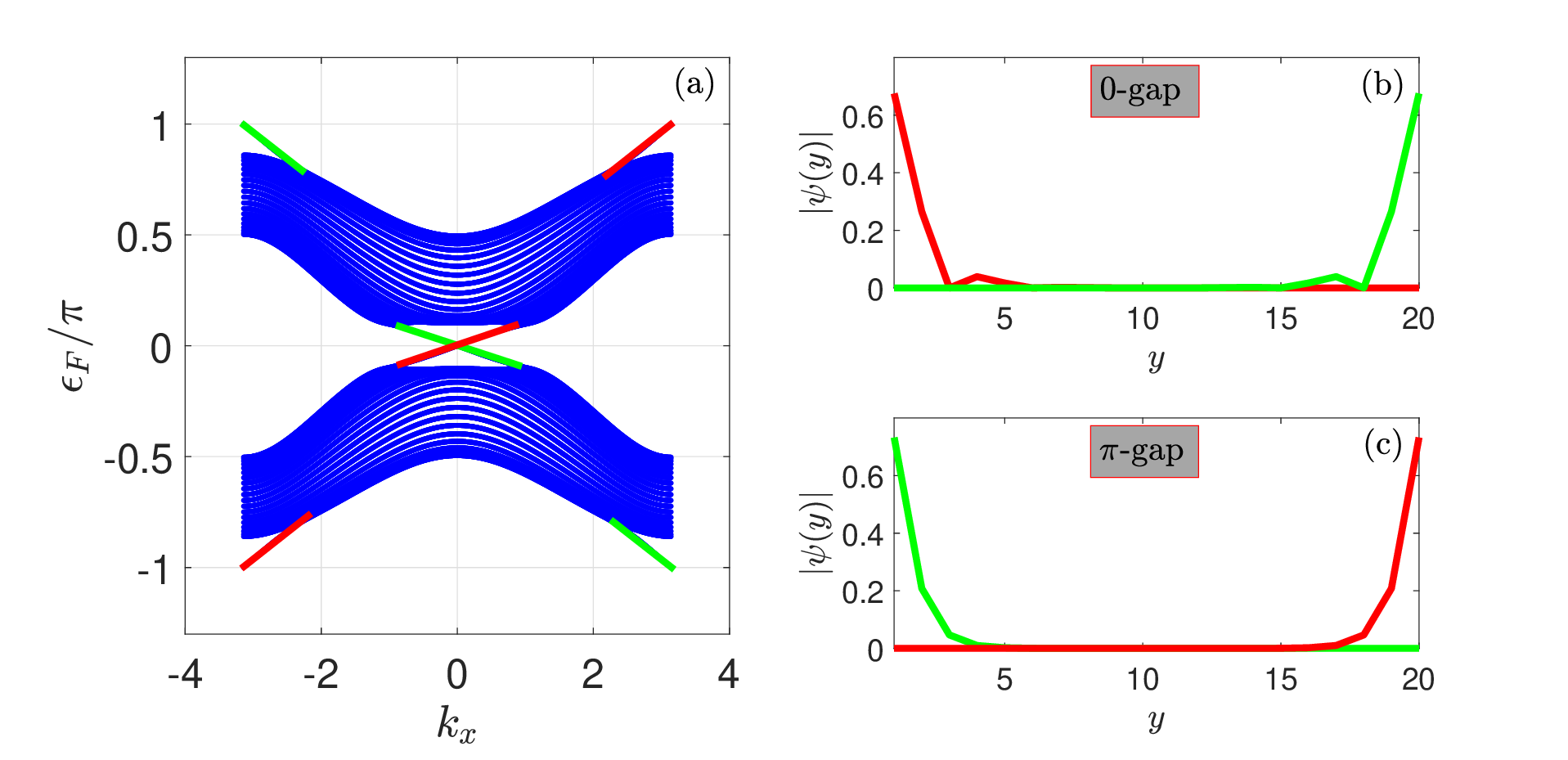}
	\caption{The quasienergy spectrum of PK-QWZ model (a), where the boundary condition along the $x$-direction is periodic, $L_y=20$, $u=1.3\pi$ and $u_0=0.2\pi$. Also presented are the chiral edge modes traversing the zero quasienergy gap (b) and the chiral edge modes traversing the $\pi$ quasienergy gap (c), where $|\psi(y)|^2 = |\psi(y)_\uparrow|^2 + |\psi(y)_\downarrow|^2$. Its obvious that the right-handed edge modes that traversing the zero quasienergy gap are localized at the bottom of system [$y=1$] (b, red solid line), while the right-handed edge modes that traversing the zero quasienergy gap are localized at the top of system [$y=L_y$] (c, red solid line), while the left-handed chiral edge modes (green solid line) are the other way around.}
	\label{spec_wave}
\end{figure}

For the brevity of discussion, we set $T=1$.
Its easy to find that
\begin{eqnarray}
	\sigma_x e^{-iH^T_F(-\mathbf{k})} \sigma_x &=& \sigma_x e^{iu\sigma_z}\sigma_x\sigma_x e^{-iH^T_{\text{qwz}}(-\mathbf{k})} \sigma_x,\nonumber\\
	 &=& e^{iu\sigma_z} e^{iH_{\text{qwz}}(\mathbf{k})},\nonumber\\
	 &=& e^{iH_F(\mathbf{k})}.
\end{eqnarray}
Consequently, the symmetry of $H_F(\mathbf{k})$ is the same as $H_{\text{qwz}}(\mathbf{k})$, that $\sigma_xH^T_F(-\mathbf{k})\sigma_x=-H_F(\mathbf{k})$, and it belongs to class D as well.
Then, the topology contained in the quasi-band of $H_F(\mathbf{k})$ can be extracted by a Chern number, called the Floquet Chern number $C_F$, in which the Berry curvature is defined within its lower quasi-band (i.e., for $\epsilon_F<0$).
According to the bulk-boundary correspondence, the presence of chiral edge modes is expected when $C_F\neq0$.
As present in Fig. \ref{spec_wave}, there are four chiral edge modes.
These edge modes form two pairs, one pair of them traversing the zero quasienergy gap, and another pair traversing the $\pi$ quasienergy gap correspondingly.
Notably, the chirality of edge modes reside in the zero quasienergy gap and reside in the $\pi$ quasienergy gap is different, that they are propagating at opposite directions.
This gives rise to the question of what the topological invariant for $H_F(\mathbf{k})$ would be.

\begin{figure*}
	\centering
	\includegraphics[width=1.0\textwidth]{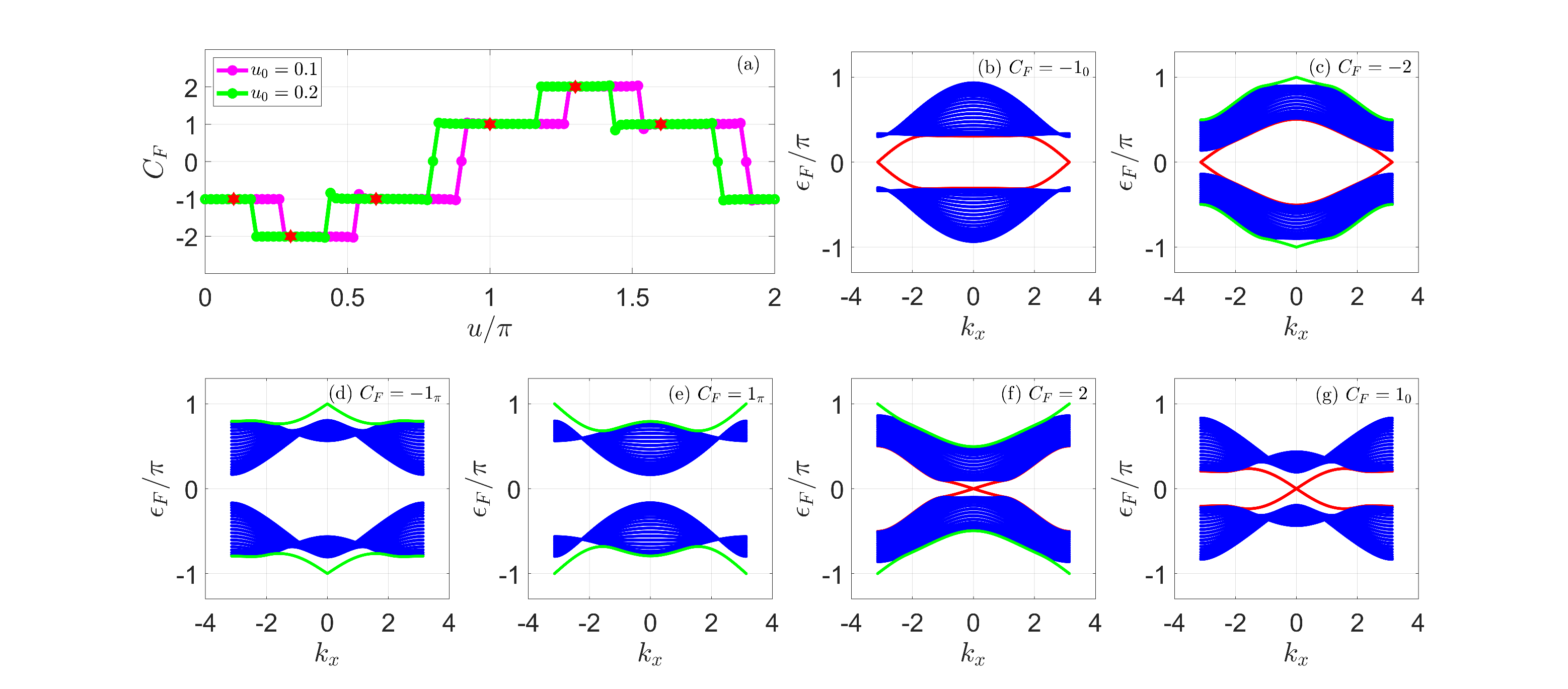}
	\caption{ (a) The Floquet Chern number $C_F$ as a function of $u$ for $u_0=0.1\pi$ and $u_0=0.2\pi$ in the PK-QWZ model. The quansi-bands of PK-QWZ model for $u=0.1\pi$ (b), $u=0.3\pi$ (c), $u=0.6\pi$ (d), $u=\pi$ (e), $u=1.3\pi$ (f) and $u=1.6\pi$ (g) separately, where $u_0=0.2\pi$, $L_y=20$. The boundary condition along the $x$-direction is periodic. For the topological phases with $C_F=\pm1$, there are two chiral edge modes, that either traversing the zero quasienergy gap (marked with red) or traversing the $\pi$ quasienergy gap (marked with green). For the topological phases with $C_F=\pm2$, there are four chiral edge modes. They are separated in two pairs, one pair of them traverse the zero quasienergy gap, and the other pair of them traverse the $\pi$ quasienergy gap.}
	\label{CF_quasi_band}
\end{figure*}

Different from the static two-band system that the Chern number uniquely defines the net number of chiral edge modes crossing the band gap, the appearance of chiral edge modes in the Floquet system might related to the case $C_F=0$.
This is due to the unbounded nature of the quasi-band, that there would have equal numbers of chiral edge modes entering the band from below and exiting above.
Consequently, their net contribution to the Chern number remains zero, as discussed in Ref. \cite{anomalous_edge_modes}.
However, this phenomena happens only if these two edge modes have the same chirality.
Fortunately, this is not true for PK-QWZ model.
The edge modes that with opposite chirality are located at the same edge, one is traversing the zero quasienergy gap and another one is traversing the $\pi$ quasienergy gap, as illustrated in Fig. \ref{spec_wave}, so the net contribution to the Chern number is nonzero.
As the results, the edge modes of the PK-QWZ model can be accurately characterized by the Chern number.

The topology of $H_F(\mathbf{k})$ is decoded in its algebraic structure, and can be formally written as
\begin{equation}\label{floquet_func}
H_F(\mathbf{k}) = \xi(\mathbf{k})\mathds{1}_{2\times2} + \mathbf{n}(\mathbf{k})\vec{\sigma}.
\end{equation}
Form BCH formula, its evident that $\xi(\mathbf{k})=0$.
The quansienery spectrum of PK-QWZ model is
\begin{equation}
	\epsilon_{\mathbf{k}}=\sqrt{\left[n^x(\mathbf{k})\right]^2+\left[n^y(\mathbf{k})\right]^2+\left[n^z(\mathbf{k})\right]^2}.
\end{equation}
The Floquet Chern number corresponds the winding number of the mapping from the BZ to the Bloch vector $\hat{\mathbf{n}}(\mathbf{k})$, defined as
\begin{equation}
	\mathbf{k}\rightarrow\hat{\mathbf{n}}(\mathbf{k})=\frac{\mathbf{n}(\mathbf{k})}{|\mathbf{n}(\mathbf{k})|}.
\end{equation}
In geometric terms, this is equivalent to counting the number of Skyrmions in the Bloch vector manifold, which can be expressed as
\begin{equation}
	C_F = \frac{1}{4\pi}\int d^2\mathbf{k}\left( \frac{\partial\hat{\mathbf{n}}(\mathbf{k})}{\partial k_x}\times \frac{\partial\hat{\mathbf{n}}(\mathbf{k})}{\partial k_y}\right) \cdot\hat{\mathbf{n}}(\mathbf{k}).
\end{equation}
And because the unit vector $\hat{\mathbf{n}}(\mathbf{k})$ is reside on a unit sphere $S^2$, then $C_F$ is equal to the times that $\hat{\mathbf{n}}(\mathbf{k})$ winds around $S^2$ as well.

Unfortunately, the exact expression of $H_F(\mathbf{k})$ is impossible by using the BCH formula (Appendix A).
Nevertheless, the value of $C_F$ can be determined numerically.
When $C_F=\pm2$, it is expected to find four chiral edge modes. These edge modes form two pairs, with one pair traversing the zero quasienergy gap and the other pair traversing the $\pi$ quasienergy gap, as depicted in Fig. \ref{CF_quasi_band} (c, f).
Surprisingly, there are two phases for $C_F=-1$, denoted as $-1_0$ and $-1_\pi$.
The edge modes traversing the zero quasienergy gap for the phase $C_F=-1_0$, while the edge modes traversing the $\pi$ quasienergy gap for the phase $C_F=-1_\pi$, see Fig. \ref{CF_quasi_band} (b, d).
And there are two phases for $C_F=1$, denoted as $1_0$ and $1_\pi$ as well, see Fig. \ref{CF_quasi_band} (e, g).

\section{The phase diagram of PK-QWZ model}
\begin{figure}[H]
	\centering
	\includegraphics[width=0.5\textwidth]{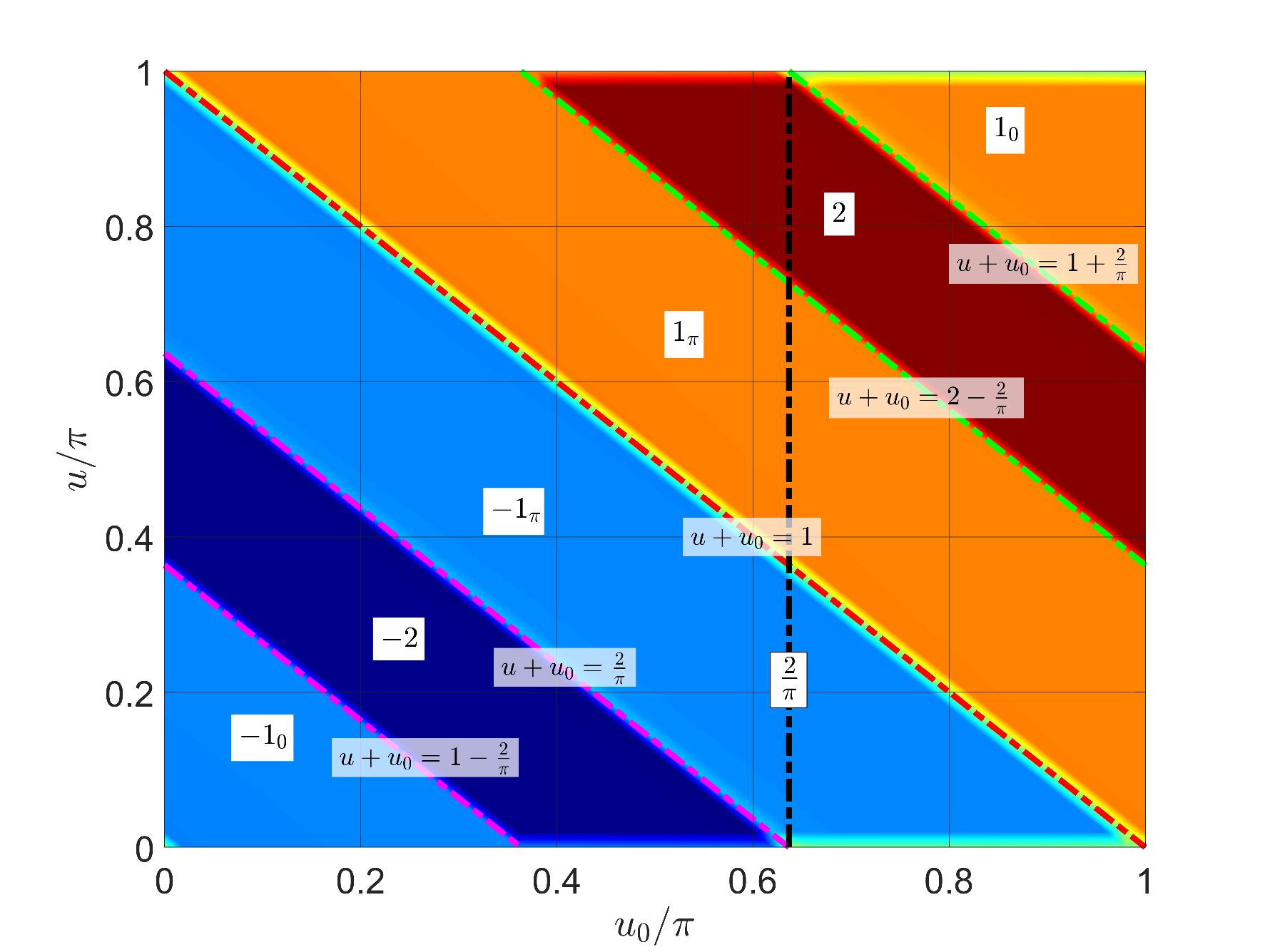}
	\caption{The phase diagram of PK-QWZ model. Where the magenta dashed lines represent the phase boundaries between the phase with $C_F=-2$ and those with $C_F=-1$, the red dashed lines are the phase boundaries between the phase with $C_F=-1_\pi$ and those with $C_F=1_\pi$, green dashed lines mark the phase boundaries between the phase with $C_F=2$ and those with $C_F=1$, and black dashed line is the critical point of static QWZ model where $u_0=2$.}
	\label{phase_diagram}
\end{figure}

The phase diagram of PK-QWZ model is depicted in Fig. \ref{phase_diagram}, illustrating the variation of the Floquet Chern number $C_F$ with respect to $u_0$ and $u$.
As previously mentioned, the Chern number of 2D system is equal to the numbers of Skyrmions ($N_{\text{sky.}}$) in the vector space manifold $\hat{\mathbf{n}}$ \cite{qwz_skyrmion}, in which the base manifold being the BZ $T^2$.
In this context, the phase $C_F=\pm1$ corresponds to the situation that the vector $\hat{\mathbf{n}}$ winds around the unit sphere $S^2$ once, while the phase $C_F=\pm2$ signifies that the vector $\hat{\mathbf{n}}$ winds around the Bloch sphere $S^2$ twice.
In other words, the phase diagram also reflects the variation of $N_{\text{sky.}}$ over the parameters $u_0$ and $u$.

The phase boundaries can be uncovered by investigating the number of Skyrmions, denoted as $N_{\text{sky.}}$, in vector space $\hat{\mathbf{n}}$.
Despite the absence of the exact expression for $H_F(\mathbf{k})$, the low-energy effective theory is sufficient for characterizing the topology of $H_F(\mathbf{k})$.
In the low energy limit, $H_F(\mathbf{k})$ reduces to a Dirac Hamiltonian with the momentum matrices $\sigma_x$ and $\sigma_y$, while the mass matrix is $\sigma_z$ (see Appendix A).
Thus, $N_{\text{sky.}}$ is entirely determined by the variation of $n^z(\mathbf{k})$ over the BZ, to be precise, $N_{\text{sky.}}$ is equal to the number of times that the sign of $n^z(\mathbf{k})$ changes across the BZ.
There are two specific points in the BZ that need consideration.
The first, denoted as $\Gamma$-type, corresponds to points $(k_x,k_y)=(0,0)$ and $(\pm\pi,\pm\pi)$; the second point, denoted as $\Lambda$-type, is located at $(0,\pm\pi)$ and $(\pm\pi,0)$.

\begin{figure}[H]
	\centering
	\includegraphics[width=.5\textwidth]{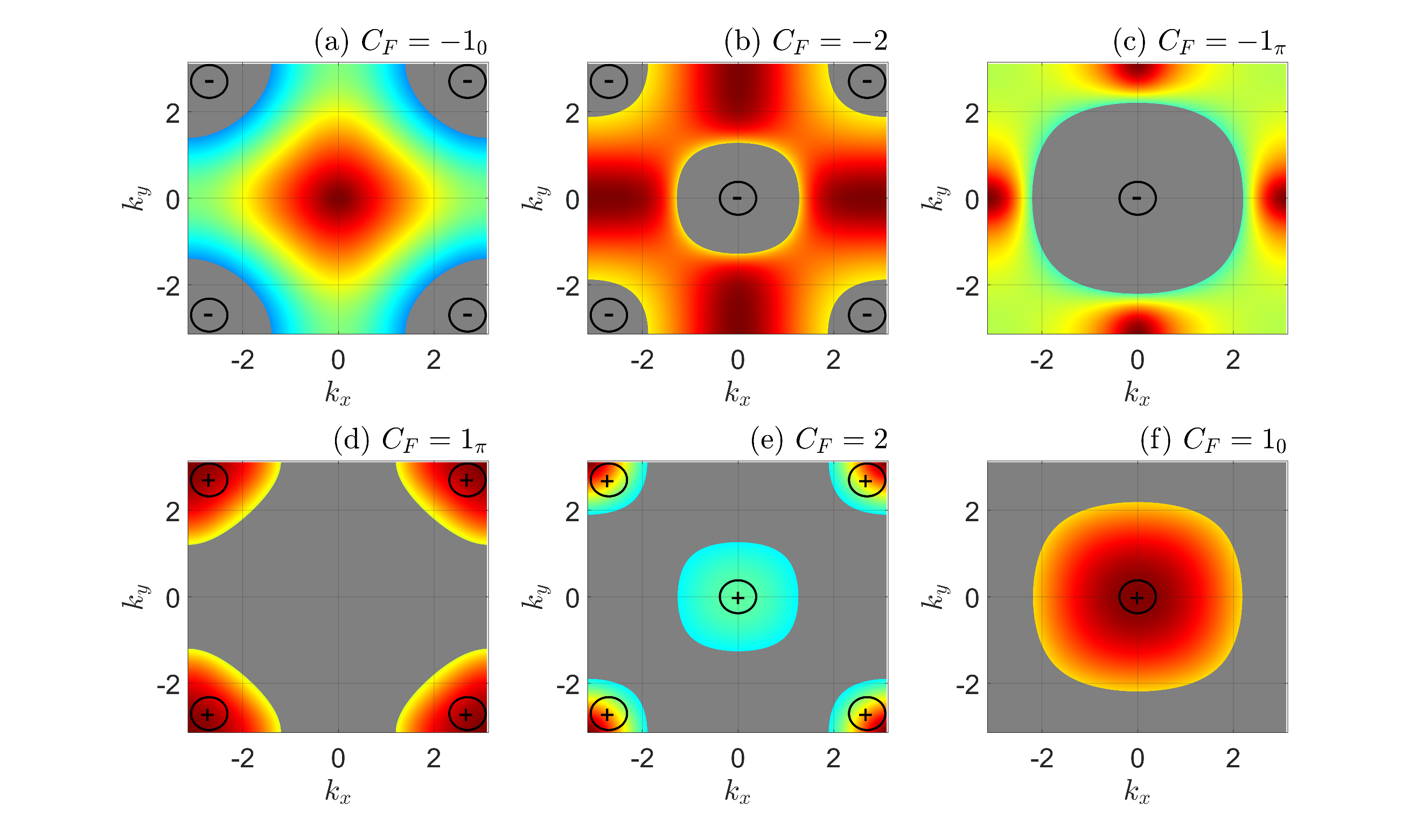}
	\caption{The contour plot of $n_z(k_x,k_y)$ for different values of $u$ with fixed $u_0=0.2\pi$, the presented cases are for $u=0.1\pi$ (a), $u=0.3\pi$ (b), $u=0.6\pi$ (c), $u=\pi$ (d), $u=1.3\pi$ (e) and $u=1.6\pi$ (f) separately. In these plots, the gray areas represent regions $(n^z)<0$. The phases $C_F=\pm1$ is for the case that the sign of $n^z$ changes only once across the BZ (a,c,d,f). While the phase with $C_F=\pm2$ is for the case that the sign of $n^z$ changes twice across the BZ (b,e).}
	\label{spec_vec_z}
\end{figure}

For $\Gamma$-type, we have
\begin{equation}
	n^z_{\Gamma,\pm} = u + u_0 \pm 2,
\end{equation}
where "$+$" is for $(0,0)$ and "$-$" is for $(\pm\pi,\pm\pi)$.
For $\Lambda$-type, we have
\begin{equation}
	n^z_\Lambda = u + u_0.
\end{equation}
As the results, $N_{\text{sky.}}$ is obtained by identifying the variation of $n^z_{\Gamma,\pm}$ and $n^z_\Lambda$ as $u_0$ and $u$ are varied.
When $-2<u+u_0<\pi-2$, it is evident that $n^z_\Lambda>0$, $n^z_{\Gamma,+}>0$, and $n^z_{\Gamma,-}<0$. In this case, the sign of $n_z$ changes only once in the BZ, leading to $N_{\text{sky.}}=1$, corresponding to the topological phase with $C_F=-1_0$. This situation is depicted in Fig. \ref{spec_vec_z} (a).
When $\pi-2<u+u_0<2$, we find that $n^z_\Lambda>0$ and $n^z_{\Gamma,\pm}<0$. In this situation, the sign of $n_z$ changes twice in the BZ, resulting $N_{\text{sky.}}=2$, which corresponds to the topological phase with $C_F=-2$, see Fig. \ref{spec_vec_z} (b).
So the phase boundary between the phase $C_F=-1_0$ and the phase $C_F=-2$ is $u+u_0+2=\pi$, as shown in Fig. \ref{phase_diagram}.
As $u+u_0$ is increases, that $2<u+u_0<\pi$, we find that $n^z_\Lambda>0$, $n^z_{\Gamma,+}<0$ and $n^z_{\Gamma,-}>0$.
In this case, the sign of $n_z$ changes only once in the BZ, leading to $N_{\text{sky.}}=1$, which corresponds to the topological phase with $C_F=-1_\pi$, see Fig. \ref{spec_vec_z} (c).
Consequently, the phase boundary between the phase $C_F=-2$ and the phase $C_F=-1_\pi$ is $u+u_0=2$, as presented in Fig. \ref{phase_diagram}.

\begin{table}[H]	
	\centering
	\setlength{\tabcolsep}{5mm}{
		\begin{tabular}{c|ccc|c}
			\hline
			\hline
			$C_F$ & $n^z_{\Gamma,+}$ & $n^z_{\Lambda}$ & $n^z_{\Gamma,-}$ & $N_{\text{sky.}}$ \\
			\hline
			$-1_0$ & $+$ & $+$ & $-$ & 1 \\
			$-2$ & $-$ & $+$ & $-$ & 2 \\
			$-1_\pi$ & $-$ & $+$ & $+$ & 1 \\
			$1_\pi$ & $-$ & $-$ & $+$ & 1 \\
			$2$ & $+$ & $-$ & $+$ & 2 \\
			$1_0$ & $+$ & $-$ & $-$ & 1 \\
			\hline
			\hline
	\end{tabular}}
	\caption{The sign of mass $n^z$ at $\Lambda$-points and $\Gamma$-points for different topological states, where "$+$" is for $n^z>0$ and "$-$" is for $n^z<0$.}
	\label{mass_val}
\end{table}

With the same approach, all configurations of $n^z_{\Gamma,\pm}$ and $n^z_\Lambda$ for different topological phases are uncovered, which is summarized in Table \ref{mass_val}.
And the phase boundaries are determined by the critical point where $n^z_{\Gamma,\pm}$ or $n^z_\Lambda$ changes sign, see Table \ref{phase_transition_data}.

\begin{table*}	
	\setlength{\tabcolsep}{4mm}{
		\begin{tabular}{c|ccc cc}
			\hline
			\hline
			Phase 1\quad\&\quad phase 2 & phase boundary & q. g. & q. g. closing point $(k_x,k_y)$&  OPT vs. $u$  & OPT vs. $u_0$ \\
			\hline
			$C_F=-1_0$\quad\&\quad$C_F=-2$ & $u+u_0+2=\pi$ & $\pi$ & $(0,0)$ & 2 & 2\\
			$C_F=-2$\quad\&\quad$C_F=-1_\pi$ & $u+u_0=2$ & $0$ & $(\pm\pi,\pm\pi)$ & 3 & 3\\
			$C_F=-1_\pi$\quad\&\quad$C_F=1_\pi$ & $u+u_0=\pi$ & $\pi$ & $(0,\pm\pi)$ and $(\pm\pi,0)$& 3 & 1\\
			$C_F=1_\pi$\quad\&\quad$C_F=2$ & $u+u_0+2=2\pi$ & $0$ & $(0,0)$ & 2 &$\smallsetminus$\\
			$C_F=2$\quad\&\quad$C_F=1_0$ & $u+u_0=2+\pi$ & $\pi$  & $(\pm\pi,\pm\pi)$ & 3 &$\smallsetminus$\\
			$C_F=1_0$\quad\&\quad$C_F=-1_0$& $u+u_0=2\pi$ & $0$ & $(0,\pm\pi)$ and $(\pm\pi,0)$ & 3 & $\smallsetminus$ \\
			\hline
			\hline
	\end{tabular}}
	\caption{The phase boundaries, the types of quasienergy gap (q. g.) that is closing, the points where quasienergy gap is closed in the BZ and the order of phase transition (OPT) respects to $u$ and $u_0$ respectively in the PK-QWZ model. Where "$\smallsetminus$" means that there isn't has well defined Floquet stationary state.}
	\label{phase_transition_data}
\end{table*}

\section{The von Neumann entropy of Floquet stationary state in PK-QWZ model}
A quantum system never completely decoupled from its environment, that the thermalization or the dissipation would finally leads the system relax to a stationary state.
However, rare is known about the stationary states of a periodically driven systems, called the Floquet stationary states. 
Related research is a very hot topic recently, for example the quantum time crystal \cite{quantum_TC,Floquet_TC}.
Furthermore, the study of the thermodynamic properties in the Floquet system is of great importance, relevant research involves to the unique Floquet-Gibbs state \cite{floquet_Gibbs_cond,floquet_Gibbs,floquet_Gibbs_chaotic,floquet_Gibbs_dissipative}, and the exceptional selection rules \cite{floquet_selection_rule}.

\begin{figure}[H]
	\centering
	\includegraphics[width=0.5\textwidth]{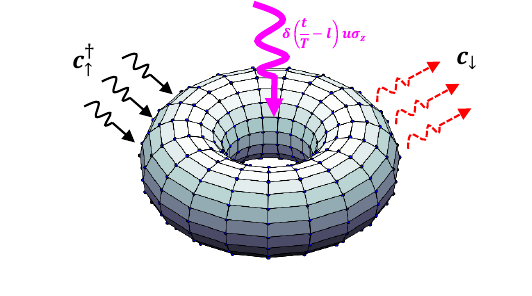}
	\caption{Pictorial illustration of PK-QWZ model that is weakly coupled to the Markovian environment, in which the occupation of particles with up-spin is boosted while the the particles with down-spin are dissipated away. And the geometry of system is a torus, that the boundary conditions are periodic both for the $x$ direction and the $y$ direction.}
	\label{grap_qwz_NE}
\end{figure}

In this section, we want to investigate the implications of Floquet topological phases within the context of open quantum systems.
This correspond to the situation that the periodically driven system is weakly coupled to a Markovian environment, allowing for the presence of a Floquet Lindbladian \cite{floquet_Lindbladian}.
In the case of a periodically kicked free fermionic system, we found that its single-particle correlation matrix of the Floquet stationary state $C^{F,s}$ satisfies a Sylvester equation (Appendix B)
\begin{equation}\label{floquet_ss}
	e^{iX_1 T}P_-  = e^{iX_1T} e^{iX_0T}C^{F,s} - C^{F,s}e^{iX_1^\dagger T}e^{iX_0^\dagger T}.
\end{equation}
Here, $X_0 = -H_0^T-i\left( M_g + M_l^T\right)$, $X_1 = -H_1^T$, and $P_-$ is related to the static stationary states $C^s$ (see Appendix B).
However, there is no reason to expects that $C^{F,s}$ reduces to $C^s$ as $u=0$. This is because that there is a time evolution factor $e^{-iX_0 T}$ in the Sylvester equation (\ref{floquet_ss}), which distinguish itself from static systems. 
We introduce the Lindblad operator as
\begin{equation}
	L^g_j = \sqrt{\gamma} c^\dagger_{j,\uparrow},\qquad L^l_j = \sqrt{\gamma} c_{j,\downarrow},
\end{equation}
where $j$ is the site index.
That due to the coupling with the environment, the occupation number of particles with up-spin is boosted while the particles with down-spin are dissipated away.
The non-Hermitian terms in $X_0$ are
\begin{equation}
	M_g = \mathds{1}_{L_x \times L_y}\otimes\begin{pmatrix}
		1&0\\
		0&0
	\end{pmatrix},\quad M_l=\mathds{1}_{L_x \times L_y}\otimes\begin{pmatrix}
		0&0\\
		0&1
	\end{pmatrix}.
\end{equation}
Meanwhile, despite the edge effects is vital for the topological phase, the bulk property of system is easier to study and can also reflects the topological properties.
So, the geometry of system is assumed as a torus with periodic boundary conditions in both the $x$ and $y$ directions, as depicted in Fig. \ref{grap_qwz_NE}.

The single-particle correlation matrix is a key quantity that has direct connections to various aspects of the system. It is related to the occupation numbers of particles, the von Neumann entropy, and the particle currents in the system. Among these, the von Neumann entropy is particularly significant as it serves as an important measure for understanding the remarkable properties of Floquet stationary states.

The von Neumann entropy is defined as $S = -\text{Tr}\left( \rho \ln \rho \right)$, which is used to quantifies the amount of quantum information that contained in a many-body quantum state. $S=0$ for a pure state, $S=1$ for the maximally mixed state, while $0<S<1$ for a general mixed state.
This information-theoretic entropy is a good candidate for thermodynamic entropy \cite{entropy_production}, that the singularities of $S$ is also a good signature of phase transition.
In a free fermionic system, the von Neumann entropy of its subsystem can be related to the eigenvalues of the corresponding single-particle correlation matrix, which is given by \cite{entropy_expression}
\begin{equation}\label{S_free_f}
	S = -\sum_{j}\left[ \xi_j\ln\xi_j+(1-\xi_j)\ln(1-\xi_j)\right],
\end{equation}
where ${\xi_j}$ are the eigenvalues of single-particle correlation matrix.
Then, by regarding $\rho$ as a subsystem of system $\rho_{\text{s+E}}$, Eq. (\ref{S_free_f}) is valid to evaluate the von Neumann entropy of $\rho$ as well.

Thus, when $C^{F,s}$ is obtained, the von Neumann entropy of Floquet stationary state is settled.
In our scenario that the geometry of system is a torus, we are interested in the average value of von Neumann entropy
\begin{equation}
	\bar{S}=\frac{1}{2L_xL_y}\frac{S}{\ln2},
\end{equation}
where $S$ is the von Neumann entropy for the Floquet stationary state, $L_x$ and $L_y$ represent the length of the torus in the $x$ and $y$ directions respectively. The factor $1/\ln2$ is included to normalize the von Neumann entropy. 

\begin{figure}[H]
	\centering
	\includegraphics[width=0.5\textwidth]{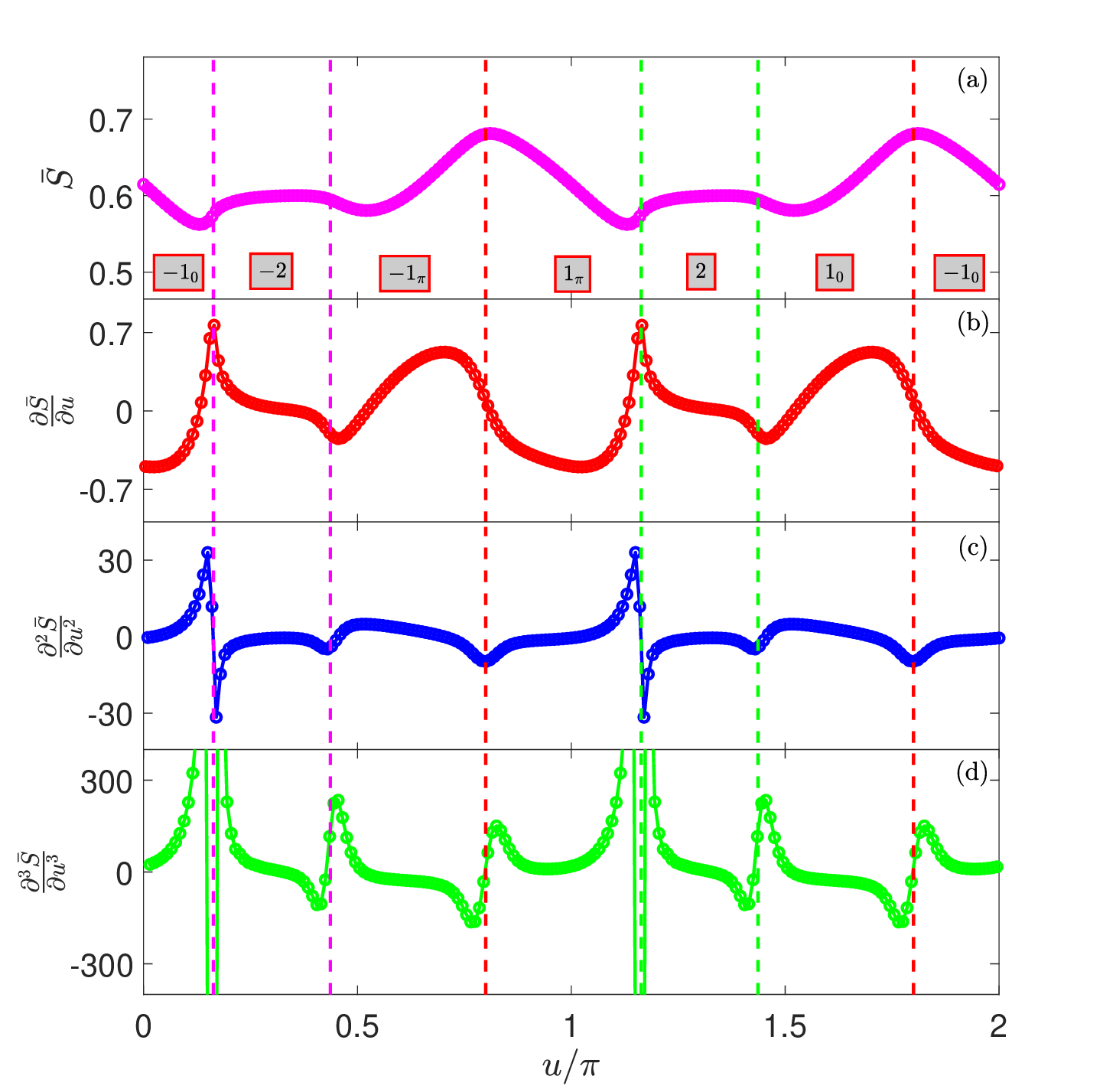}
	\caption{The average von Neumann entropy $\bar{S}$ (a) of Floquet stationary state in the torus-like PK-QWZ model, and the first-order (b), the second-order (c) and the third-order (d) derivatives of $\bar{S}$ with respects to $u$. Where $u_0=0.2\pi$, $\gamma=0.002$, and $L_x=L_y=30$. It worth noting that $\bar{S}$ is periodic in $u$ with a period of $\pi$.}
	\label{FS_curr_ent_u}
\end{figure}

In Fig. \ref{FS_curr_ent_u}, we present the dependence of average von Neumann entropy $\bar{S}$ of Floquet stationary state on the parameter $u$, where $u_0$ is fixed.
Notably, we observe a periodic behavior of $\bar{S}$ as a function of $u$, with this periodicity having a period of $\pi$. This periodic pattern is a direct consequence of the periodic nature of the Sylvester equation in Eq. (\ref{floquet_ss}) respect to $u$, which possesses an intrinsic $\pi$-periodicity.
Furthermore, within the topological phase characterized by $C_F=\pm2$, $\bar{S}$ features a plateau. And $\bar{S}$ reach the maximum value precisely at the critical points marking the transition between phases with $C_F=\pm1_{0,\pi}$ and those with $C_F=\mp1_{0,\pi}$, as presented in Fig. \ref{FS_curr_ent_u} (a).
Remarkably, the discontinuities that emerge in the derivatives of $\bar{S}$ respect to $u$ are precisely coincide with critical points separating different topological phases.
We find that $\partial_u^2\bar{S}$ is discontinuous both for the critical point that separating the phases with $C_F=-1_0$ and $C_F=-2$ and for the critical point between the phases with $C_F=1_\pi$ and $C_F=2$, which means that they are the second-order phase transitions respect to $u$.
Additionally, $\partial_u^3\bar{S}$ is discontinuities at the critical point that separating the phases with $C_F=-1_\pi$ and $C_F=-2$, the critical point between the phases with $C_F=1_0$ and $C_F=2$, as well as the critical point between the phases with $C_F=\pm1_{0,\pi}$ and $C_F=\mp1_{0,\pi}$, these imply that they are third-order phase transitions respect to $u$.
A comprehensive summary of these transitions can be found in Table \ref{phase_transition_data}.

\begin{figure}[H]
	\centering
	\includegraphics[width=0.5\textwidth]{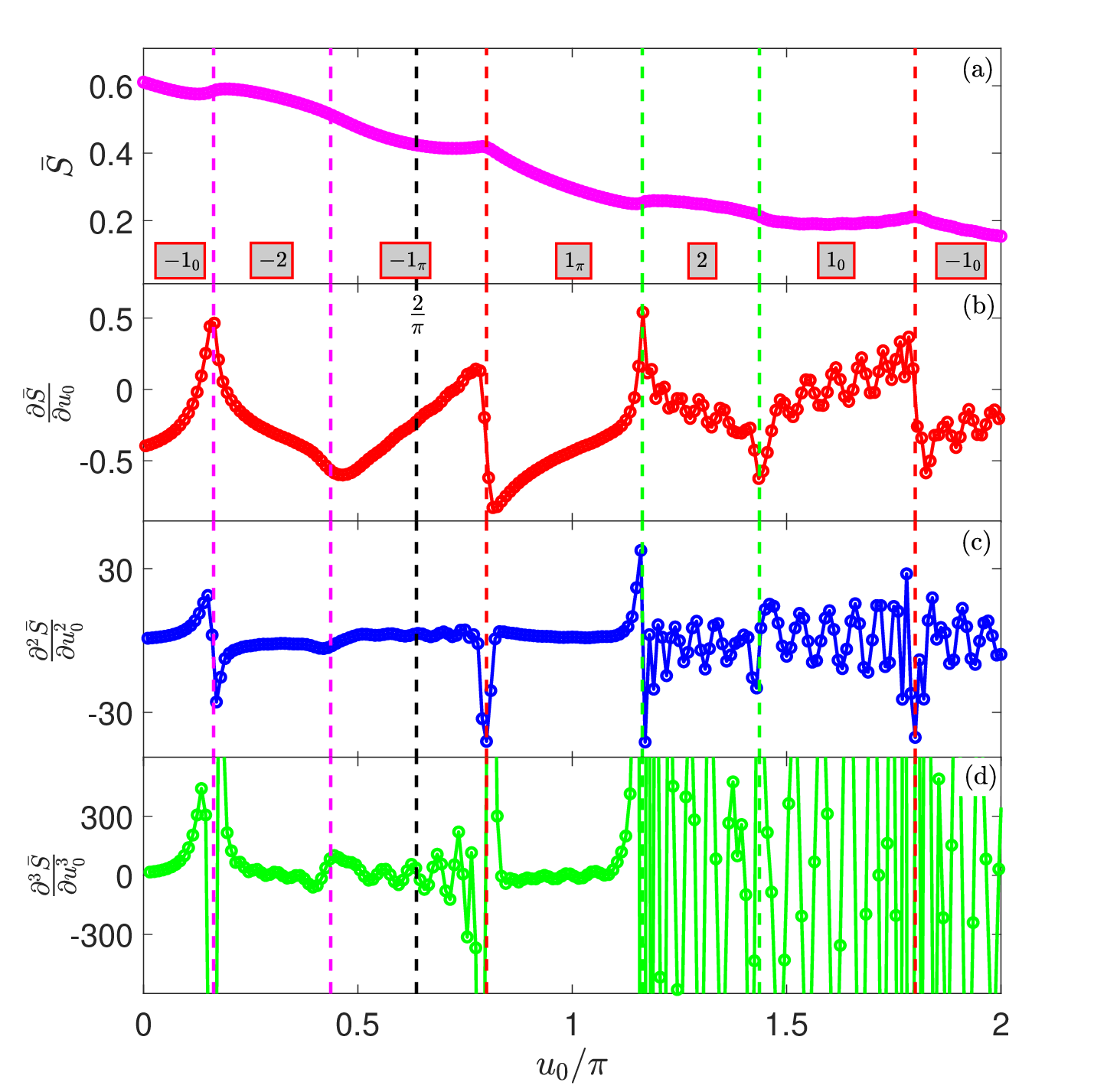}
	\caption{The average von Neumann entropy $\bar{S}$ (a) of Floquet stationary state in the torus-like PK-QWZ model, and the first-order (b), the second-order (c) and the third-order (d) derivatives of $\bar{S}$ with respect to $u_0$. Where $u=0.2\pi$, $\gamma=0.002$, and $L_x=L_y=30$. Moreover, $\bar{S}$ decreases as $u_0$ increases, and well-defined phases only exist when $u_0<2\pi-2-u$, because the derivative of $\bar{S}$ with respect to $u_0$ becomes chaotic beyond this limit.}
	\label{FS_curr_ent_u0}
\end{figure}

The value of $\bar{S}$ in Floquet stationary states versus $u_0$ is studied as well, where $u$ is fixed.
The results reveal a decreasing trend of $\bar{S}$ as $u_0$ increases, as depicted in Fig. \ref{FS_curr_ent_u0} (a).
Notably, it is observed that the phase is well-defined only if $u_0\leq2\pi-2-u$, otherwise the derivation of $\bar{S}$ with respect to $u_0$ exhibits chaotic behavior.
We find that $\partial_{u_0}\bar{S}$ exhibits a discontinuity at the critical point between the phases with $C_F=-1_\pi$ and $C_F=1_\pi$, indicating a first-order phase transition with respects to $u_0$, see Fig. \ref{FS_curr_ent_u0} (b).
Additionally, $\partial^2_{u_0}\bar{S}$ exhibits a discontinuity at the critical point between the phases with $C_F=-1_0$ and $C_F=-2$, marking a second-order phase transition respects $u_0$, see Fig. \ref{FS_curr_ent_u0} (c).
Finally, $\partial^3_{u_0}\bar{S}$ exhibits a discontinuity at the critical point between the phases with $C_F=-2$ and $C_F=-1_\pi$, indicating a third-order phase transition with respect to $u_0$, see Fig. \ref{FS_curr_ent_u0} (d).
These findings are summarized in Table \ref{phase_transition_data}.

\section{Conclusion and Discussion}
The massive Dirac Hamiltonian is vital in determining the topology of the various models, with the classification space corresponds to a set of permissible Dirac mass terms \cite{Clliford_classification}.
Therefore, when the Dirac mass term is periodically kicked, the Floquet Chern insulators can inherit the symmetry class of the corresponding static Hamiltonian, which can be seen in Appendix C.
As the symmetry class remains unchanged, the topological phases in the Floquet system can be characterized with the same topological invariant.
However, because the quasienergy band is unbounded, the value of topological numbers might be different, and so is the number of protected edge modes.
Fortunately, due to the inconsistency of chirality between the edge modes traversing the zero quasienergy gap and those traversing the $\pi$ quasienergy gap, the bulk-boundary correspondence remains valid in this system.
In our illustrative example, the PK-QWZ model, the Floquet Chern number $C_F$ is used to characterize the corresponding topological phases.
We found that there are six topological phases in total, denoted as $C_F=\{-1_0,-2,-1_\pi,1_\pi,2,1_0\}$.
Moreover, the phase diagram is established by quantifying the numbers of Skyrmions in the Floquet Bloch vector space.

Additionally, analogous to the Lyapunov function of static stationary states, a Sylvester equation (\ref{floquet_ss}) of Floquet stationary states is derived in this paper.
This equation is valid as the Floquet Lindbladian can be applied \cite{floquet_Lindbladian}.
By studying the von Neumann entropy of Floquet stationary states in the PK-QWZ model, we unveil the orders of phase transitions between different topological phases, which are summarized in Table \ref{phase_transition_data}.

One step forward, the invariance of symmetry class in Floquet Chern insulators whose mass term is periodically kicked opens up avenues for considering higher values of topological numbers in various topological systems, such as the $\mathbb{Z}_2$ TIs and topological superconductors.
This prospect is our future research direction.

\appendix

\section{The approximate expression of Floquet operator}
The Euler equation directly leads to
\begin{equation}
	\exp(i\phi\hat{\mathbf{n}}\cdot\vec{\sigma}) = \cos \phi + i\hat{\mathbf{n}}\cdot\vec{\sigma} \sin \phi.
\end{equation}
By express the Hamiltonian and the Floquet operator as
\begin{eqnarray}
	H_0 &=& E_\mathbf{k}\left( r^x \sigma_x + r^y \sigma_y + r^z \sigma_z\right),\\
	H_F &=& \epsilon_\mathbf{k} \left( \hat{n}^x \sigma_x + \hat{n}^y \sigma_y + \hat{n}^z \sigma_z\right),
\end{eqnarray}
where $
	r^x = \frac{\sin k_x}{E_\mathbf{k}}, \, r^y = \frac{\sin k_y}{E_\mathbf{k}}, \, r^z = \frac{u_0 + \cos k_x +\cos k_y}{E_\mathbf{k}}
$,
and $\left( \hat{n}^x\right)^2 + \left( \hat{n}^y\right)^2 + \left( \hat{n}^z\right)^2 =1$. The energy of QWZ model is
\begin{eqnarray}
	E_\mathbf{k} = \sqrt{\sin^2 k_x + \sin^2 k_y + \left( u_0 + \cos k_x +\cos k_y \right)^2 },\qquad
\end{eqnarray}
and $\epsilon_{\mathbf{k}}$ is the quasienergy of system, that $\epsilon_{\mathbf{k}}\in[-\pi,\pi)$.
Then, the time evolution operator in the PK-QWZ model can be rewritten as
\begin{widetext}	
\begin{eqnarray}
	U &=& \exp\left[ -iH_0(\mathbf{k}) \right] \exp\left( -iH_1 \right),\nonumber\\
	&=& \left[ \cos\left( E_\mathbf{k} \right) - i \left( r^x \sigma_x + r^y \sigma_y + r^z \sigma_z\right)\sin\left( E_\mathbf{k} \right) \right] \left(  \cos u - i\sigma_z \sin u \right),\nonumber\\
	&=& \left[ \cos\left( E_\mathbf{k} \right)\cos u + r^z\sin\left( E_\mathbf{k} \right)\sin u  \right] - i\sigma_z \left[ \cos\left( E_\mathbf{k} \right) \sin u + r^z\sin\left( E_\mathbf{k} \right) \cos u \right]\nonumber\\
	&& -i\sigma_x \sin\left( E_\mathbf{k} \right) \left[ r^x \cos u + r^y \sin u \right] - i\sigma_y \sin\left( E_\mathbf{k} \right) \left[ r^y \cos u - r^x \sin u \right].
\end{eqnarray}
\end{widetext}
For simplicity, $U(T=1)$ has been denoted as $U$.
Furthermore, an equivalent expression is $U=\exp\left( -iH_F \right)$, so
\begin{equation}
	U = \cos \epsilon_\mathbf{k} -i \left( \hat{n}^x \sigma_x + \hat{n}^y \sigma_y + \hat{n}^z \sigma_z\right)\sin \epsilon_\mathbf{k} .
\end{equation}
Thus, we have the following equations
\begin{eqnarray}
	\cos \epsilon_\mathbf{k} &=& \cos\left( E_\mathbf{k} \right)\cos u + r^z\sin\left( E_\mathbf{k} \right)\sin u, \\
	\hat{n}^x\sin \epsilon_\mathbf{k} &=& \sin\left( E_\mathbf{k} \right) \left[ r^x \cos u + r^y \sin u \right], \\
	\hat{n}^y\sin \epsilon_\mathbf{k} &=& \sin\left( E_\mathbf{k} \right) \left[ r^y \cos u - r^x \sin u \right],  \\
	\hat{n}^z\sin \epsilon_\mathbf{k} &=&  \cos\left( E_\mathbf{k} \right) \sin u + r^z\sin\left( E_\mathbf{k} \right) \cos u .\qquad
\end{eqnarray}

In the low-energy limit that $E_\mathbf{k}\rightarrow0_+$ as  $k_x\rightarrow0_+$ and $k_y\rightarrow0_+$, we have
\begin{eqnarray}
	\cos \epsilon_\mathbf{k} &\simeq& \cos u + \left(u_0 + 2\right)\sin u, \\
	\hat{n}^x\sin \epsilon_\mathbf{k} &\simeq&  k_x \cos u + k_y \sin u , \\
	\hat{n}^y\sin \epsilon_\mathbf{k} &\simeq&  k_y \cos u - k_x \sin u ,  \\
	\hat{n}^z\sin \epsilon_\mathbf{k} &\simeq&  \sin u + \left(u_0 + 2\right) \cos u .\qquad
\end{eqnarray}

Meanwhile, as the energy gap is closed, the zero quasienergy gap might be closed ($\epsilon_{\mathbf{k}}\rightarrow0_+$) as $u\rightarrow0_+$, that we have
\begin{eqnarray}
	n^x &\simeq& k_x + u k_y , \\
	n^y  &\simeq& k_y - u k_x ,  \\
	n^z &\simeq& u + u_0 + 2 .\qquad
\end{eqnarray}

As the results, when the periodic driving is the Dirac mass term of Chern insulators, the algebraic structure of effective low-energy theory of Floquet operator is identical to its static counterpart.

However, the approximate formula of Floquet operator would provides more insights about the effective low-energy physics.
In which the Baker-Campbell-Hausdorff formula is used, that
\begin{eqnarray}
	e^{A} e^{B} &=& \exp\left[ A+B+\frac{1}{2}[A,B]+\frac{1}{12} [A,[A,B]] +  \right. \nonumber\\
	&&\left.+\frac{1}{12}[B,[B,A]]+\frac{1}{24}[B,[A,[A,B]]]\dots \right].\qquad
\end{eqnarray}
Then, we considering the following commutators
\begin{widetext}	
\begin{eqnarray}
	i[ -iH_{\text{qwz}}(\mathbf{k}), -iu\sigma_z] &=& -2u\sin k_x \sigma_y + 2u \sin k_y \sigma_x,\\	
	i\left[ -iu\sigma_z, -[ u\sigma_z, H_{\text{qwz}} (\mathbf{k})] \right] &=& 4u^2 \sin k_x \sigma_x + 4u^2 \sin k_y \sigma_y,\\
	i\left[ -iH_{\text{qwz}}(\mathbf{k}), -[ H_{\text{qwz}} (\mathbf{k}), u\sigma_z] \right] &=& -4u\left(\sin^2 k_x+\sin^2 k_y\right)\sigma_z+ 4u\sin k_x\left(u_0+\cos k_x + \cos k_y\right)\sigma_x\nonumber\\
	&&+ 4u\sin k_y\left(u_0+\cos k_x + \cos k_y\right)\sigma_y,\\
	i\left[-iu\sigma_z, \left[ -iH_{\text{qwz}}(\mathbf{k}), -[ H_{\text{qwz}} (\mathbf{k}), u\sigma_z] \right]  \right] &=& 8u^2\sin k_x (u_0+\cos k_x + \cos k_y)\sigma_y \nonumber\\
	&& - 8u^2\sin k_y (u_0+\cos k_x + \cos k_y)\sigma_x.
\end{eqnarray}
\end{widetext}
Ignoring the high-order terms, we have
\begin{widetext}
\begin{eqnarray}
	n^x &\simeq&  u\sin k_y\left[1-\frac{u}{3}\left( u_0+\cos k_x + \cos k_y \right)    \right]  + \frac{u\sin k_x}{3}\left( \frac{3}{u} + u + u_0+\cos k_x + \cos k_y \right), \\
	n^y &\simeq&  -u\sin k_x\left[1-\frac{u}{3}\left( u_0+\cos k_x + \cos k_y \right)    \right] + \frac{u\sin k_y}{3}\left( \frac{3}{u} + u + u_0+\cos k_x + \cos k_y \right)  ,\\
	n^z &\simeq&  u + u_0+\cos k_x + \cos k_y - \frac{u}{3}\left(\sin^2 k_x+\sin^2 k_y\right).\qquad
\end{eqnarray}
\end{widetext}
As expected, the Floquet operator $H_F$ reduces to a Dirac Hamiltonian in the low-energy limit, which the momentum matrices are $\sigma_x$ and $\sigma_y$, and the mass matrix is $\sigma_z$.

So, there are two kinds of points in the BZ where the quasienergy gap might close, the $\Gamma$ type and the $\Lambda$ type.
If the quasienergy gap is closing at $(k_x,k_y)=(\pm\pi,\pm\pi)$ and $(k_x,k_y)=(0,0)$, the quasienergy gap is
\begin{equation}
	n^z_{\Gamma,\pm} = u+u_0\pm2,
\end{equation}
where "$+$" is for $(0,0)$, while "$-$" is for $(\pm\pi,\pm\pi)$.
And if the quasienergy gap is closing at $(k_x,k_y)=(\pm\pi,0)$ and $(k_x,k_y)=(0,\pm\pi)$, the quasienergy gap is
\begin{equation}
	n^z_{\Lambda} = u+u_0.
\end{equation}
By identifying the configuration of $n^z$, the topological property of $H_F$ is settled, as demonstrated in the main text.

\section{The Sylvester equation of periodically kicked free fermionic system}
In this section, we considering the periodically driven system is coupled to a Markovian environment. That the evolution of states is described with a Floquet Lindbladian
\begin{eqnarray}\label{Floquet_Master_eq}
	\frac{d}{dt}\rho = -\mathrm{i}\left[H(t),\rho\right]+\sum_\mu\left(2L_\mu^\dagger \rho L_\mu -\{L_\mu^\dagger L_\mu, \rho\}\right).
\end{eqnarray}
Floquet Lindbaldian is a generalization of Lindbald master equation, and it is valid when the driving strength is sufficient strong \cite{floquet_Lindbladian}.
In which $L_\mu$ corresponds to the gain/loss of particles due to the coupling to the environment.
The general form of Lindblad operator is
\begin{equation}
	L_j^g = \sum_sD^g_{j,s}c^\dagger_{j,s},\qquad L_j^l = \sum_sD^l_{j,s}c_{\mu,s},
\end{equation}
where $j$ is the site index, and $s$ corresponds to the internal degrees of freedom for each lattice site.

Due to the Gaussian quadratic property of free fremionic systems, we can use the single-particle correlation matrix to characterize the dynamics of system \cite{chiral_damping,open_qs_correlation_matrix}
\begin{equation}\label{Time_evolve}
	i\frac{d}{dt}C(t) =X(t)C(t) - C(t) X(t)^\dagger + 2\mathrm{i}M_g,
\end{equation}
where $C_{mn} = \langle c^\dagger_m c_n \rangle$. In which $X(t)$ is the time-dependent damping matrix
\begin{equation}\label{damping_m}
	X(t)=-H^T(t)-i\left( M_g + M_l^T\right),
\end{equation}
where the non-Hermitian terms $M_g$ and $M_l$ are induced by the coupling with environment, that
\begin{equation}
	\left( M_g\right)_{ij} = \sum_{\mu}D^{g*}_{\mu i}D^g_{\mu j}, \quad \left( M_l\right)_{ij} = \sum_{\mu}D^{l*}_{\mu i}D^l_{\mu j}.
\end{equation}

Unfortunately, because the Hamiltonian is time-dependent, then there is no reason to require that $\partial_t C^{F,s} = 0$, where $C^{F,s}$ is the stationary state of Floquet open quantum system.
However, its natural to require that $C^{F,s}$ is periodic in time when the Hamiltonian is periodic in time, i.e. $C^{F,s}(t+T) = C^{F,s}(t)$.

Then, we introducing the matrix \cite{floquet_stationary}
\begin{equation}
	\Xi = \begin{pmatrix}
		X&-2iM_g\\
		0&X^\dagger
	\end{pmatrix}, \qquad D = \begin{pmatrix}
		\mathds{1}&C\\
		0&0
	\end{pmatrix},
\end{equation}
the Eq. (\ref{Time_evolve}) can be rewritten as
\begin{equation}
	i\frac{d}{dt} D = [\Xi,D].
\end{equation}
There is a formal solution
\begin{eqnarray}
	D(t) &=& \mathcal{T}\exp\left( -i\int_{0}^{t} d\tau \Xi(\tau)\right)D_0\mathcal{T}\exp\left( i\int_{0}^{t} d\tau \Xi(\tau)\right),\nonumber\\
	&=& \mathcal{U}^F(t) D^s \mathcal{U}^F(-t),
\end{eqnarray}
where $\mathcal{T}$ is the time-ordering operator.
Then, the time evolution operator in a full period $T$ is
\begin{equation}
 	\mathcal{U}^F(T) = e^{-i\Xi_0 T}e^{-i\Xi_1 T},
\end{equation}
where 
\begin{equation}
	\Xi_0 = \begin{pmatrix}
		X_0 & -2iM_g\\
		0 & X_0^\dagger
	\end{pmatrix},\qquad\Xi_1=\begin{pmatrix}
	 	X_1 & 0\\
	 	0 &X_1^\dagger
	\end{pmatrix},
\end{equation}
and $X_0=-H_0^T-i(M_g+M_l^T)$, $X_1=-H_1^T$.

Furthermore, $\Xi_0$ can be rewritten as
\begin{equation}
	\Xi_0 = \begin{pmatrix}
		\mathds{1} & -C^s\\
		0 &\mathds{1}
	\end{pmatrix}\begin{pmatrix}
		X_0 & 0\\
		0 & X_0^\dagger
	\end{pmatrix}\begin{pmatrix}
		\mathds{1} & C^s\\
		0 &\mathds{1}
	\end{pmatrix},
\end{equation}
where $C^s$ satisfy the Lyapunov equation
\begin{equation}
	X_0C^s-C^sX_0^\dagger=-2iM_g.
\end{equation}
In which $C^s$ is the single-particle correlation matrix of static stationary states.
After simple algebraic derivation, we find that
\begin{eqnarray}
	\label{forward_t}
	\mathcal{U}^F(T) &=& \begin{pmatrix}
		e^{-iX_0T}e^{-iX_1T} & P_+ e^{-iX_1^\dagger T} \\
		0 & e^{-iX_0^\dagger T}e^{-iX_1^\dagger T}
	\end{pmatrix},\\
	\label{backward_t}
	\mathcal{U}^F(-T) &=& \begin{pmatrix}
		e^{iX_1T}e^{iX_0T} & e^{iX_1 T}P_-  \\
		0 & e^{iX_1^\dagger T}e^{iX_0^\dagger T}
	\end{pmatrix},\\
	P_+ &=& e^{-iX_0T} C^s - C^s e^{-iX_0^\dagger T},\\
	P_- &=& e^{iX_0T} C^s - C^s e^{iX_0^\dagger T}.
\end{eqnarray}
Finally, if there is a Floquet stationary state, it should be periodic in time with period $T$, i.e.
\begin{equation}\label{flquet_constraint}
	D^s = \mathcal{U}^F(T) D^s \mathcal{U}^F(-T).
\end{equation}
Then, substituting Eqs. (\ref{forward_t}, \ref{backward_t}) into Eq. (\ref{flquet_constraint}), we have a Sylvester equation
\begin{equation}
	e^{iX_1 T}P_-  = e^{iX_1T} e^{iX_0T}C^{F,s} - C^{F,s}e^{iX_1^\dagger T}e^{iX_0^\dagger T},
\end{equation}
where $C^{F,s}$ is the single-particle correlation matrix of Floquet stationary state.

\section{The symmetry class of Floquet operator}
In a periodically kicked system, the algebraic structure of time evolution operator is very special, in which we can proof that the symmetry class of Floquet operator is identical to its static counterpart.
Take a massive Dirac Hamiltonian as an example
\begin{equation}
	H_{D} = \sum_{l=1}^{d} k_l \gamma_l + m_0 \gamma_0,
\end{equation}
where $\{\gamma_l,\gamma_j\} = 2\delta_{lj}$ satisfy a Clliford algebra, and $d$ is the spatial dimension of system.
And $\gamma_0$ is the mass matrix that constraint by the symmetry of system.
So, the driven of periodically kicked Chern insulators is $H_1 = m \gamma_0$.

The time reversal symmetry (TRS), particle hole symmetry (PHS) and sub-lattice symmetry (SLS) of $H_{D}$ are defined as
\begin{eqnarray}
	\text{TRS}:&&\qquad U_t \cdot H_D^* \cdot U_t^{-1} = H_D,\\
	\text{PHS}:&&\qquad U_c \cdot H_D^T \cdot U_c^{-1} = -H_D,\\
	\text{SLS}:&&\qquad U_s \cdot H_D \cdot U_s^{-1} = -H_D.
\end{eqnarray}
where $U_{t,c,s}$ are unitary operators.

Then, when the TRS is satisfied for $H_D$, it directly leads to
\begin{eqnarray}
	U_t e^{iH_F^*} U_t^{-1} &=& U_t e^{iH_D^*}U_t^{-1} U_t e^{iH_1^*} U_t^{-1},\nonumber\\
	&=& e^{iH_D} e^{iH_1},\nonumber\\
	&=& e^{iH_F},
\end{eqnarray}
we have $U_t H_F^* U_t^{-1}=H_F$, the TRS is satisfied for $H_F$.

And if PHS is satisfied for $H_D$, then
\begin{eqnarray}
	U_c e^{-iH_F^T} U_c^{-1} &=& U_c e^{-iH_1^T} U_c^{-1} U_c e^{-iH_D^T}U_c^{-1} ,\nonumber\\
	&=& e^{iH_1}e^{iH_D} ,\nonumber\\
	&=& e^{iH_F},
\end{eqnarray}
we have $U_c H_F^T U_c^{-1}=-H_F$, the PHS is satisfied for $H_F$.

Finally, if SLS is satisfied for $H_D$, we have
\begin{eqnarray}
	U_s e^{-iH_F} U_s^{-1} &=& U_s e^{-iH_D}U_s^{-1} U_s e^{-iH_1} U_s^{-1},\nonumber\\
	&=& e^{iH_D}e^{iH_1} ,\nonumber\\
	&=& e^{iH_F},
\end{eqnarray}
we have $U_s H_F U_s^{-1}=-H_F$, the SLS is satisfied for $H_F$.

In conclusion, for a periodically kicked Chern insulator, the symmetry class of $H_F$ is identical to its static counterpart if the periodically driving term is the Dirac mass term.

\bibliography{references}

\begin{thebibliography}{33}%
\makeatletter
\providecommand \@ifxundefined [1]{%
 \@ifx{#1\undefined}
}%
\providecommand \@ifnum [1]{%
 \ifnum #1\expandafter \@firstoftwo
 \else \expandafter \@secondoftwo
 \fi
}%
\providecommand \@ifx [1]{%
 \ifx #1\expandafter \@firstoftwo
 \else \expandafter \@secondoftwo
 \fi
}%
\providecommand \natexlab [1]{#1}%
\providecommand \enquote  [1]{``#1''}%
\providecommand \bibnamefont  [1]{#1}%
\providecommand \bibfnamefont [1]{#1}%
\providecommand \citenamefont [1]{#1}%
\providecommand \href@noop [0]{\@secondoftwo}%
\providecommand \href [0]{\begingroup \@sanitize@url \@href}%
\providecommand \@href[1]{\@@startlink{#1}\@@href}%
\providecommand \@@href[1]{\endgroup#1\@@endlink}%
\providecommand \@sanitize@url [0]{\catcode `\\12\catcode `\$12\catcode
  `\&12\catcode `\#12\catcode `\^12\catcode `\_12\catcode `\%12\relax}%
\providecommand \@@startlink[1]{}%
\providecommand \@@endlink[0]{}%
\providecommand \url  [0]{\begingroup\@sanitize@url \@url }%
\providecommand \@url [1]{\endgroup\@href {#1}{\urlprefix }}%
\providecommand \urlprefix  [0]{URL }%
\providecommand \Eprint [0]{\href }%
\providecommand \doibase [0]{http://dx.doi.org/}%
\providecommand \selectlanguage [0]{\@gobble}%
\providecommand \bibinfo  [0]{\@secondoftwo}%
\providecommand \bibfield  [0]{\@secondoftwo}%
\providecommand \translation [1]{[#1]}%
\providecommand \BibitemOpen [0]{}%
\providecommand \bibitemStop [0]{}%
\providecommand \bibitemNoStop [0]{.\EOS\space}%
\providecommand \EOS [0]{\spacefactor3000\relax}%
\providecommand \BibitemShut  [1]{\csname bibitem#1\endcsname}%
\let\auto@bib@innerbib\@empty
\bibitem [{\citenamefont {Qi}\ and\ \citenamefont
  {Zhang}(2011)}]{review_TI_TS}%
  \BibitemOpen
  \bibfield  {author} {\bibinfo {author} {\bibfnamefont {X.-L.}\ \bibnamefont
  {Qi}}\ and\ \bibinfo {author} {\bibfnamefont {S.-C.}\ \bibnamefont {Zhang}},\
  }\href {\doibase 10.1103/RevModPhys.83.1057} {\bibfield  {journal} {\bibinfo
  {journal} {Rev. Mod. Phys.}\ }\textbf {\bibinfo {volume} {83}},\ \bibinfo
  {pages} {1057} (\bibinfo {year} {2011})}\BibitemShut {NoStop}%
\bibitem [{\citenamefont {Cayssol}\ \emph {et~al.}(2013)\citenamefont
  {Cayssol}, \citenamefont {D\'{o}ra}, \citenamefont {Simon},\ and\
  \citenamefont {Moessner}}]{floquet_TIs_review}%
  \BibitemOpen
  \bibfield  {author} {\bibinfo {author} {\bibfnamefont {J.}~\bibnamefont
  {Cayssol}}, \bibinfo {author} {\bibfnamefont {B.}~\bibnamefont {D\'{o}ra}},
  \bibinfo {author} {\bibfnamefont {F.}~\bibnamefont {Simon}}, \ and\ \bibinfo
  {author} {\bibfnamefont {R.}~\bibnamefont {Moessner}},\ }\href {\doibase
  https://doi.org/10.1002/pssr.201206451} {\bibfield  {journal} {\bibinfo
  {journal} {physica status solidi (RRL) – Rapid Research Letters}\ }\textbf
  {\bibinfo {volume} {7}},\ \bibinfo {pages} {101} (\bibinfo {year}
  {2013})}\BibitemShut {NoStop}%
\bibitem [{\citenamefont {Titum}\ \emph {et~al.}(2016)\citenamefont {Titum},
  \citenamefont {Berg}, \citenamefont {Rudner}, \citenamefont {Refael},\ and\
  \citenamefont {Lindner}}]{floquet_anderson}%
  \BibitemOpen
  \bibfield  {author} {\bibinfo {author} {\bibfnamefont {P.}~\bibnamefont
  {Titum}}, \bibinfo {author} {\bibfnamefont {E.}~\bibnamefont {Berg}},
  \bibinfo {author} {\bibfnamefont {M.~S.}\ \bibnamefont {Rudner}}, \bibinfo
  {author} {\bibfnamefont {G.}~\bibnamefont {Refael}}, \ and\ \bibinfo {author}
  {\bibfnamefont {N.~H.}\ \bibnamefont {Lindner}},\ }\href {\doibase
  10.1103/PhysRevX.6.021013} {\bibfield  {journal} {\bibinfo  {journal} {Phys.
  Rev. X}\ }\textbf {\bibinfo {volume} {6}},\ \bibinfo {pages} {021013}
  (\bibinfo {year} {2016})}\BibitemShut {NoStop}%
\bibitem [{\citenamefont {Grushin}\ \emph {et~al.}(2014)\citenamefont
  {Grushin}, \citenamefont {G\'omez-Le\'on},\ and\ \citenamefont
  {Neupert}}]{floquet_fractional_chern}%
  \BibitemOpen
  \bibfield  {author} {\bibinfo {author} {\bibfnamefont {A.~G.}\ \bibnamefont
  {Grushin}}, \bibinfo {author} {\bibfnamefont {A.}~\bibnamefont
  {G\'omez-Le\'on}}, \ and\ \bibinfo {author} {\bibfnamefont {T.}~\bibnamefont
  {Neupert}},\ }\href {\doibase 10.1103/PhysRevLett.112.156801} {\bibfield
  {journal} {\bibinfo  {journal} {Phys. Rev. Lett.}\ }\textbf {\bibinfo
  {volume} {112}},\ \bibinfo {pages} {156801} (\bibinfo {year}
  {2014})}\BibitemShut {NoStop}%
\bibitem [{\citenamefont {Rudner}\ \emph {et~al.}(2013)\citenamefont {Rudner},
  \citenamefont {Lindner}, \citenamefont {Berg},\ and\ \citenamefont
  {Levin}}]{anomalous_edge_modes}%
  \BibitemOpen
  \bibfield  {author} {\bibinfo {author} {\bibfnamefont {M.~S.}\ \bibnamefont
  {Rudner}}, \bibinfo {author} {\bibfnamefont {N.~H.}\ \bibnamefont {Lindner}},
  \bibinfo {author} {\bibfnamefont {E.}~\bibnamefont {Berg}}, \ and\ \bibinfo
  {author} {\bibfnamefont {M.}~\bibnamefont {Levin}},\ }\href {\doibase
  10.1103/PhysRevX.3.031005} {\bibfield  {journal} {\bibinfo  {journal} {Phys.
  Rev. X}\ }\textbf {\bibinfo {volume} {3}},\ \bibinfo {pages} {031005}
  (\bibinfo {year} {2013})}\BibitemShut {NoStop}%
\bibitem [{\citenamefont {Zhou}\ and\ \citenamefont
  {Zhang}(2023)}]{floquet_NH}%
  \BibitemOpen
  \bibfield  {author} {\bibinfo {author} {\bibfnamefont {L.}~\bibnamefont
  {Zhou}}\ and\ \bibinfo {author} {\bibfnamefont {D.-J.}\ \bibnamefont
  {Zhang}},\ }\href {\doibase 10.3390/e25101401} {\bibfield  {journal}
  {\bibinfo  {journal} {Entropy}\ }\textbf {\bibinfo {volume} {25}} (\bibinfo
  {year} {2023}),\ 10.3390/e25101401}\BibitemShut {NoStop}%
\bibitem [{\citenamefont {Wu}\ \emph {et~al.}(2021)\citenamefont {Wu},
  \citenamefont {Wang},\ and\ \citenamefont {An}}]{floquet_NH_higher_TIs}%
  \BibitemOpen
  \bibfield  {author} {\bibinfo {author} {\bibfnamefont {H.}~\bibnamefont
  {Wu}}, \bibinfo {author} {\bibfnamefont {B.-Q.}\ \bibnamefont {Wang}}, \ and\
  \bibinfo {author} {\bibfnamefont {J.-H.}\ \bibnamefont {An}},\ }\href
  {\doibase 10.1103/PhysRevB.103.L041115} {\bibfield  {journal} {\bibinfo
  {journal} {Phys. Rev. B}\ }\textbf {\bibinfo {volume} {103}},\ \bibinfo
  {pages} {L041115} (\bibinfo {year} {2021})}\BibitemShut {NoStop}%
\bibitem [{\citenamefont {Pan}\ and\ \citenamefont
  {Zhou}(2020)}]{floquet_NH_second_TIs}%
  \BibitemOpen
  \bibfield  {author} {\bibinfo {author} {\bibfnamefont {J.}~\bibnamefont
  {Pan}}\ and\ \bibinfo {author} {\bibfnamefont {L.}~\bibnamefont {Zhou}},\
  }\href {\doibase 10.1103/PhysRevB.102.094305} {\bibfield  {journal} {\bibinfo
   {journal} {Phys. Rev. B}\ }\textbf {\bibinfo {volume} {102}},\ \bibinfo
  {pages} {094305} (\bibinfo {year} {2020})}\BibitemShut {NoStop}%
\bibitem [{\citenamefont {Zhou}(2020)}]{floquet_NH_maj}%
  \BibitemOpen
  \bibfield  {author} {\bibinfo {author} {\bibfnamefont {L.}~\bibnamefont
  {Zhou}},\ }\href {\doibase 10.1103/PhysRevB.101.014306} {\bibfield  {journal}
  {\bibinfo  {journal} {Phys. Rev. B}\ }\textbf {\bibinfo {volume} {101}},\
  \bibinfo {pages} {014306} (\bibinfo {year} {2020})}\BibitemShut {NoStop}%
\bibitem [{\citenamefont {Zhang}\ and\ \citenamefont
  {Gong}(2020)}]{floquet_NH_topo_phase}%
  \BibitemOpen
  \bibfield  {author} {\bibinfo {author} {\bibfnamefont {X.}~\bibnamefont
  {Zhang}}\ and\ \bibinfo {author} {\bibfnamefont {J.}~\bibnamefont {Gong}},\
  }\href {\doibase 10.1103/PhysRevB.101.045415} {\bibfield  {journal} {\bibinfo
   {journal} {Phys. Rev. B}\ }\textbf {\bibinfo {volume} {101}},\ \bibinfo
  {pages} {045415} (\bibinfo {year} {2020})}\BibitemShut {NoStop}%
\bibitem [{\citenamefont {Ke}\ \emph {et~al.}(2023)\citenamefont {Ke},
  \citenamefont {Wen}, \citenamefont {Zhao},\ and\ \citenamefont
  {Wang}}]{floquet_NH_skin}%
  \BibitemOpen
  \bibfield  {author} {\bibinfo {author} {\bibfnamefont {S.}~\bibnamefont
  {Ke}}, \bibinfo {author} {\bibfnamefont {W.}~\bibnamefont {Wen}}, \bibinfo
  {author} {\bibfnamefont {D.}~\bibnamefont {Zhao}}, \ and\ \bibinfo {author}
  {\bibfnamefont {Y.}~\bibnamefont {Wang}},\ }\href {\doibase
  10.1103/PhysRevA.107.053508} {\bibfield  {journal} {\bibinfo  {journal}
  {Phys. Rev. A}\ }\textbf {\bibinfo {volume} {107}},\ \bibinfo {pages}
  {053508} (\bibinfo {year} {2023})}\BibitemShut {NoStop}%
\bibitem [{\citenamefont {Roy}\ and\ \citenamefont
  {Harper}(2017)}]{floquet_periodic_table}%
  \BibitemOpen
  \bibfield  {author} {\bibinfo {author} {\bibfnamefont {R.}~\bibnamefont
  {Roy}}\ and\ \bibinfo {author} {\bibfnamefont {F.}~\bibnamefont {Harper}},\
  }\href {\doibase 10.1103/PhysRevB.96.155118} {\bibfield  {journal} {\bibinfo
  {journal} {Phys. Rev. B}\ }\textbf {\bibinfo {volume} {96}},\ \bibinfo
  {pages} {155118} (\bibinfo {year} {2017})}\BibitemShut {NoStop}%
\bibitem [{\citenamefont {Kitagawa}\ \emph {et~al.}(2010)\citenamefont
  {Kitagawa}, \citenamefont {Berg}, \citenamefont {Rudner},\ and\ \citenamefont
  {Demler}}]{topo_cha_periodical_driven}%
  \BibitemOpen
  \bibfield  {author} {\bibinfo {author} {\bibfnamefont {T.}~\bibnamefont
  {Kitagawa}}, \bibinfo {author} {\bibfnamefont {E.}~\bibnamefont {Berg}},
  \bibinfo {author} {\bibfnamefont {M.}~\bibnamefont {Rudner}}, \ and\ \bibinfo
  {author} {\bibfnamefont {E.}~\bibnamefont {Demler}},\ }\href {\doibase
  10.1103/PhysRevB.82.235114} {\bibfield  {journal} {\bibinfo  {journal} {Phys.
  Rev. B}\ }\textbf {\bibinfo {volume} {82}},\ \bibinfo {pages} {235114}
  (\bibinfo {year} {2010})}\BibitemShut {NoStop}%
\bibitem [{\citenamefont {Nathan}\ and\ \citenamefont
  {Rudner}(2015)}]{floquet_Bloch}%
  \BibitemOpen
  \bibfield  {author} {\bibinfo {author} {\bibfnamefont {F.}~\bibnamefont
  {Nathan}}\ and\ \bibinfo {author} {\bibfnamefont {M.~S.}\ \bibnamefont
  {Rudner}},\ }\href {\doibase 10.1088/1367-2630/17/12/125014} {\bibfield
  {journal} {\bibinfo  {journal} {New Journal of Physics}\ }\textbf {\bibinfo
  {volume} {17}},\ \bibinfo {pages} {125014} (\bibinfo {year}
  {2015})}\BibitemShut {NoStop}%
\bibitem [{\citenamefont {Wang}\ \emph {et~al.}(2013)\citenamefont {Wang},
  \citenamefont {Steinberg}, \citenamefont {Jarillo-Herrero},\ and\
  \citenamefont {Gedik}}]{floquet_Bloch_observation}%
  \BibitemOpen
  \bibfield  {author} {\bibinfo {author} {\bibfnamefont {Y.~H.}\ \bibnamefont
  {Wang}}, \bibinfo {author} {\bibfnamefont {H.}~\bibnamefont {Steinberg}},
  \bibinfo {author} {\bibfnamefont {P.}~\bibnamefont {Jarillo-Herrero}}, \ and\
  \bibinfo {author} {\bibfnamefont {N.}~\bibnamefont {Gedik}},\ }\href
  {\doibase 10.1126/science.1239834} {\bibfield  {journal} {\bibinfo  {journal}
  {Science}\ }\textbf {\bibinfo {volume} {342}},\ \bibinfo {pages} {453}
  (\bibinfo {year} {2013})}\BibitemShut {NoStop}%
\bibitem [{\citenamefont {Asb\'oth}\ \emph {et~al.}(2014)\citenamefont
  {Asb\'oth}, \citenamefont {Tarasinski},\ and\ \citenamefont
  {Delplace}}]{zero_pi_topo_num}%
  \BibitemOpen
  \bibfield  {author} {\bibinfo {author} {\bibfnamefont {J.~K.}\ \bibnamefont
  {Asb\'oth}}, \bibinfo {author} {\bibfnamefont {B.}~\bibnamefont
  {Tarasinski}}, \ and\ \bibinfo {author} {\bibfnamefont {P.}~\bibnamefont
  {Delplace}},\ }\href {\doibase 10.1103/PhysRevB.90.125143} {\bibfield
  {journal} {\bibinfo  {journal} {Phys. Rev. B}\ }\textbf {\bibinfo {volume}
  {90}},\ \bibinfo {pages} {125143} (\bibinfo {year} {2014})}\BibitemShut
  {NoStop}%
\bibitem [{\citenamefont {Shirley}(1965)}]{floquet_theorem}%
  \BibitemOpen
  \bibfield  {author} {\bibinfo {author} {\bibfnamefont {J.~H.}\ \bibnamefont
  {Shirley}},\ }\href {\doibase 10.1103/PhysRev.138.B979} {\bibfield  {journal}
  {\bibinfo  {journal} {Phys. Rev.}\ }\textbf {\bibinfo {volume} {138}},\
  \bibinfo {pages} {B979} (\bibinfo {year} {1965})}\BibitemShut {NoStop}%
\bibitem [{\citenamefont {Zhou}\ and\ \citenamefont
  {Gong}(2018)}]{double_kicked_rotor}%
  \BibitemOpen
  \bibfield  {author} {\bibinfo {author} {\bibfnamefont {L.}~\bibnamefont
  {Zhou}}\ and\ \bibinfo {author} {\bibfnamefont {J.}~\bibnamefont {Gong}},\
  }\href {\doibase 10.1103/PhysRevA.97.063603} {\bibfield  {journal} {\bibinfo
  {journal} {Phys. Rev. A}\ }\textbf {\bibinfo {volume} {97}},\ \bibinfo
  {pages} {063603} (\bibinfo {year} {2018})}\BibitemShut {NoStop}%
\bibitem [{\citenamefont {Morimoto}\ and\ \citenamefont
  {Furusaki}(2013)}]{Clliford_classification}%
  \BibitemOpen
  \bibfield  {author} {\bibinfo {author} {\bibfnamefont {T.}~\bibnamefont
  {Morimoto}}\ and\ \bibinfo {author} {\bibfnamefont {A.}~\bibnamefont
  {Furusaki}},\ }\href {\doibase 10.1103/PhysRevB.88.125129} {\bibfield
  {journal} {\bibinfo  {journal} {Phys. Rev. B}\ }\textbf {\bibinfo {volume}
  {88}},\ \bibinfo {pages} {125129} (\bibinfo {year} {2013})}\BibitemShut
  {NoStop}%
\bibitem [{\citenamefont {Qi}\ \emph {et~al.}(2006)\citenamefont {Qi},
  \citenamefont {Wu},\ and\ \citenamefont {Zhang}}]{qwz_skyrmion}%
  \BibitemOpen
  \bibfield  {author} {\bibinfo {author} {\bibfnamefont {X.-L.}\ \bibnamefont
  {Qi}}, \bibinfo {author} {\bibfnamefont {Y.-S.}\ \bibnamefont {Wu}}, \ and\
  \bibinfo {author} {\bibfnamefont {S.-C.}\ \bibnamefont {Zhang}},\ }\href
  {\doibase 10.1103/PhysRevB.74.085308} {\bibfield  {journal} {\bibinfo
  {journal} {Phys. Rev. B}\ }\textbf {\bibinfo {volume} {74}},\ \bibinfo
  {pages} {085308} (\bibinfo {year} {2006})}\BibitemShut {NoStop}%
\bibitem [{\citenamefont {Wilczek}(2012)}]{quantum_TC}%
  \BibitemOpen
  \bibfield  {author} {\bibinfo {author} {\bibfnamefont {F.}~\bibnamefont
  {Wilczek}},\ }\href {\doibase 10.1103/PhysRevLett.109.160401} {\bibfield
  {journal} {\bibinfo  {journal} {Phys. Rev. Lett.}\ }\textbf {\bibinfo
  {volume} {109}},\ \bibinfo {pages} {160401} (\bibinfo {year}
  {2012})}\BibitemShut {NoStop}%
\bibitem [{\citenamefont {Else}\ \emph {et~al.}(2016)\citenamefont {Else},
  \citenamefont {Bauer},\ and\ \citenamefont {Nayak}}]{Floquet_TC}%
  \BibitemOpen
  \bibfield  {author} {\bibinfo {author} {\bibfnamefont {D.~V.}\ \bibnamefont
  {Else}}, \bibinfo {author} {\bibfnamefont {B.}~\bibnamefont {Bauer}}, \ and\
  \bibinfo {author} {\bibfnamefont {C.}~\bibnamefont {Nayak}},\ }\href
  {\doibase 10.1103/PhysRevLett.117.090402} {\bibfield  {journal} {\bibinfo
  {journal} {Phys. Rev. Lett.}\ }\textbf {\bibinfo {volume} {117}},\ \bibinfo
  {pages} {090402} (\bibinfo {year} {2016})}\BibitemShut {NoStop}%
\bibitem [{\citenamefont {Shirai}\ \emph {et~al.}(2015)\citenamefont {Shirai},
  \citenamefont {Mori},\ and\ \citenamefont {Miyashita}}]{floquet_Gibbs_cond}%
  \BibitemOpen
  \bibfield  {author} {\bibinfo {author} {\bibfnamefont {T.}~\bibnamefont
  {Shirai}}, \bibinfo {author} {\bibfnamefont {T.}~\bibnamefont {Mori}}, \ and\
  \bibinfo {author} {\bibfnamefont {S.}~\bibnamefont {Miyashita}},\ }\href
  {\doibase 10.1103/PhysRevE.91.030101} {\bibfield  {journal} {\bibinfo
  {journal} {Phys. Rev. E}\ }\textbf {\bibinfo {volume} {91}},\ \bibinfo
  {pages} {030101} (\bibinfo {year} {2015})}\BibitemShut {NoStop}%
\bibitem [{\citenamefont {Engelhardt}\ \emph {et~al.}(2019)\citenamefont
  {Engelhardt}, \citenamefont {Platero},\ and\ \citenamefont
  {Cao}}]{floquet_Gibbs}%
  \BibitemOpen
  \bibfield  {author} {\bibinfo {author} {\bibfnamefont {G.}~\bibnamefont
  {Engelhardt}}, \bibinfo {author} {\bibfnamefont {G.}~\bibnamefont {Platero}},
  \ and\ \bibinfo {author} {\bibfnamefont {J.}~\bibnamefont {Cao}},\ }\href
  {\doibase 10.1103/PhysRevLett.123.120602} {\bibfield  {journal} {\bibinfo
  {journal} {Phys. Rev. Lett.}\ }\textbf {\bibinfo {volume} {123}},\ \bibinfo
  {pages} {120602} (\bibinfo {year} {2019})}\BibitemShut {NoStop}%
\bibitem [{\citenamefont {Ketzmerick}\ and\ \citenamefont
  {Wustmann}(2010)}]{floquet_Gibbs_chaotic}%
  \BibitemOpen
  \bibfield  {author} {\bibinfo {author} {\bibfnamefont {R.}~\bibnamefont
  {Ketzmerick}}\ and\ \bibinfo {author} {\bibfnamefont {W.}~\bibnamefont
  {Wustmann}},\ }\href {\doibase 10.1103/PhysRevE.82.021114} {\bibfield
  {journal} {\bibinfo  {journal} {Phys. Rev. E}\ }\textbf {\bibinfo {volume}
  {82}},\ \bibinfo {pages} {021114} (\bibinfo {year} {2010})}\BibitemShut
  {NoStop}%
\bibitem [{\citenamefont {Shirai}\ \emph {et~al.}(2016)\citenamefont {Shirai},
  \citenamefont {Thingna}, \citenamefont {Mori}, \citenamefont {Denisov},
  \citenamefont {Hänggi},\ and\ \citenamefont
  {Miyashita}}]{floquet_Gibbs_dissipative}%
  \BibitemOpen
  \bibfield  {author} {\bibinfo {author} {\bibfnamefont {T.}~\bibnamefont
  {Shirai}}, \bibinfo {author} {\bibfnamefont {J.}~\bibnamefont {Thingna}},
  \bibinfo {author} {\bibfnamefont {T.}~\bibnamefont {Mori}}, \bibinfo {author}
  {\bibfnamefont {S.}~\bibnamefont {Denisov}}, \bibinfo {author} {\bibfnamefont
  {P.}~\bibnamefont {Hänggi}}, \ and\ \bibinfo {author} {\bibfnamefont
  {S.}~\bibnamefont {Miyashita}},\ }\href {\doibase
  10.1088/1367-2630/18/5/053008} {\bibfield  {journal} {\bibinfo  {journal}
  {New Journal of Physics}\ }\textbf {\bibinfo {volume} {18}},\ \bibinfo
  {pages} {053008} (\bibinfo {year} {2016})}\BibitemShut {NoStop}%
\bibitem [{\citenamefont {Engelhardt}\ and\ \citenamefont
  {Cao}(2021)}]{floquet_selection_rule}%
  \BibitemOpen
  \bibfield  {author} {\bibinfo {author} {\bibfnamefont {G.}~\bibnamefont
  {Engelhardt}}\ and\ \bibinfo {author} {\bibfnamefont {J.}~\bibnamefont
  {Cao}},\ }\href {\doibase 10.1103/PhysRevLett.126.090601} {\bibfield
  {journal} {\bibinfo  {journal} {Phys. Rev. Lett.}\ }\textbf {\bibinfo
  {volume} {126}},\ \bibinfo {pages} {090601} (\bibinfo {year}
  {2021})}\BibitemShut {NoStop}%
\bibitem [{\citenamefont {Schnell}\ \emph {et~al.}(2020)\citenamefont
  {Schnell}, \citenamefont {Eckardt},\ and\ \citenamefont
  {Denisov}}]{floquet_Lindbladian}%
  \BibitemOpen
  \bibfield  {author} {\bibinfo {author} {\bibfnamefont {A.}~\bibnamefont
  {Schnell}}, \bibinfo {author} {\bibfnamefont {A.}~\bibnamefont {Eckardt}}, \
  and\ \bibinfo {author} {\bibfnamefont {S.}~\bibnamefont {Denisov}},\ }\href
  {\doibase 10.1103/PhysRevB.101.100301} {\bibfield  {journal} {\bibinfo
  {journal} {Phys. Rev. B}\ }\textbf {\bibinfo {volume} {101}},\ \bibinfo
  {pages} {100301} (\bibinfo {year} {2020})}\BibitemShut {NoStop}%
\bibitem [{\citenamefont {Landi}\ and\ \citenamefont
  {Paternostro}(2021)}]{entropy_production}%
  \BibitemOpen
  \bibfield  {author} {\bibinfo {author} {\bibfnamefont {G.~T.}\ \bibnamefont
  {Landi}}\ and\ \bibinfo {author} {\bibfnamefont {M.}~\bibnamefont
  {Paternostro}},\ }\href {\doibase 10.1103/RevModPhys.93.035008} {\bibfield
  {journal} {\bibinfo  {journal} {Rev. Mod. Phys.}\ }\textbf {\bibinfo {volume}
  {93}},\ \bibinfo {pages} {035008} (\bibinfo {year} {2021})}\BibitemShut
  {NoStop}%
\bibitem [{\citenamefont {Chang}\ \emph {et~al.}(2020)\citenamefont {Chang},
  \citenamefont {You}, \citenamefont {Wen},\ and\ \citenamefont
  {Ryu}}]{entropy_expression}%
  \BibitemOpen
  \bibfield  {author} {\bibinfo {author} {\bibfnamefont {P.-Y.}\ \bibnamefont
  {Chang}}, \bibinfo {author} {\bibfnamefont {J.-S.}\ \bibnamefont {You}},
  \bibinfo {author} {\bibfnamefont {X.}~\bibnamefont {Wen}}, \ and\ \bibinfo
  {author} {\bibfnamefont {S.}~\bibnamefont {Ryu}},\ }\href {\doibase
  10.1103/PhysRevResearch.2.033069} {\bibfield  {journal} {\bibinfo  {journal}
  {Phys. Rev. Res.}\ }\textbf {\bibinfo {volume} {2}},\ \bibinfo {pages}
  {033069} (\bibinfo {year} {2020})}\BibitemShut {NoStop}%
\bibitem [{\citenamefont {Song}\ \emph {et~al.}(2019)\citenamefont {Song},
  \citenamefont {Yao},\ and\ \citenamefont {Wang}}]{chiral_damping}%
  \BibitemOpen
  \bibfield  {author} {\bibinfo {author} {\bibfnamefont {F.}~\bibnamefont
  {Song}}, \bibinfo {author} {\bibfnamefont {S.}~\bibnamefont {Yao}}, \ and\
  \bibinfo {author} {\bibfnamefont {Z.}~\bibnamefont {Wang}},\ }\href {\doibase
  10.1103/PhysRevLett.123.170401} {\bibfield  {journal} {\bibinfo  {journal}
  {Phys. Rev. Lett.}\ }\textbf {\bibinfo {volume} {123}},\ \bibinfo {pages}
  {170401} (\bibinfo {year} {2019})}\BibitemShut {NoStop}%
\bibitem [{\citenamefont {Asadian}\ \emph {et~al.}(2013)\citenamefont
  {Asadian}, \citenamefont {Manzano}, \citenamefont {Tiersch},\ and\
  \citenamefont {Briegel}}]{open_qs_correlation_matrix}%
  \BibitemOpen
  \bibfield  {author} {\bibinfo {author} {\bibfnamefont {A.}~\bibnamefont
  {Asadian}}, \bibinfo {author} {\bibfnamefont {D.}~\bibnamefont {Manzano}},
  \bibinfo {author} {\bibfnamefont {M.}~\bibnamefont {Tiersch}}, \ and\
  \bibinfo {author} {\bibfnamefont {H.~J.}\ \bibnamefont {Briegel}},\ }\href
  {\doibase 10.1103/PhysRevE.87.012109} {\bibfield  {journal} {\bibinfo
  {journal} {Phys. Rev. E}\ }\textbf {\bibinfo {volume} {87}},\ \bibinfo
  {pages} {012109} (\bibinfo {year} {2013})}\BibitemShut {NoStop}%
\bibitem [{\citenamefont {van Caspel}\ \emph {et~al.}(2018)\citenamefont {van
  Caspel}, \citenamefont {Arze},\ and\ \citenamefont
  {Castillo}}]{floquet_stationary}%
  \BibitemOpen
  \bibfield  {author} {\bibinfo {author} {\bibfnamefont {M.}~\bibnamefont {van
  Caspel}}, \bibinfo {author} {\bibfnamefont {S.~E.~T.}\ \bibnamefont {Arze}},
  \ and\ \bibinfo {author} {\bibfnamefont {I.~P.}\ \bibnamefont {Castillo}},\
  }\href {https://api.semanticscholar.org/CorpusID:119355892} {\bibfield
  {journal} {\bibinfo  {journal} {SciPost Physics}\ } (\bibinfo {year}
  {2018})}\BibitemShut {NoStop}%
\end{thebibliography}%

\end{document}